\documentclass[prd,aps,preprintnumbers,floats,floatfix,superscriptaddress,preprintnumbers,showpacs,eqsecnum,nofootinbib,notitlepage,twocolumn]{revtex4-1} 
\usepackage[utf8]{inputenc}
\usepackage{latexsym,array,theorem,mathrsfs,bm,float}
\usepackage{psfrag}
\usepackage{subcaption}
\usepackage{amsfonts,amsmath,amssymb,latexsym,array,afterpage,theorem,mathrsfs,bm,float,epsfig,color,graphicx,tabularx,here,multirow}

\def\eqm#1{\begin{eqnarray*} #1\end{eqnarray*}}
\def\eqn#1{\begin{eqnarray} #1\end{eqnarray}}
\def\d{\partial}
\def\l{\left}\def\r{\right}

\def\cf#1{\mathcal #1}

\def\={&=&}

\newcommand{\nn}{\nonumber \\}
\newcommand{\bea}{\begin{eqnarray}}
\newcommand{\ena}{\end{eqnarray}}
\newcommand{\beann}{\begin{eqnarray*}}
\newcommand{\enann}{\end{eqnarray*}}
\newcommand{\sGamma}{\scalebox{.5}{$\Gamma$}}
\newcommand{\tsGamma}{\scalebox{.5}{$\tilde{\Gamma}$}}
\newcommand{\Gnabla}{\stackrel{\,\sGamma\,}{\nabla}}
\newcommand{\Gsquare}{\stackrel{\,\sGamma\,}{\square}}
\newcommand{\tGsquare}{\stackrel{\,\tsGamma\,}{\square}}
\newcommand{\GR}{\stackrel{\,\sGamma\,}{R}}

\newcommand{\fsquare}{\stackrel{\,\circ\,}{\square}}
\newcommand{\vect}[1]{\,\!\!\!\mbox{ \boldmath $#1$}}

\newcommand{\gsim}{\, \mbox{\raisebox{-1.ex}
{$\stackrel{\textstyle>}{\textstyle\sim}$}}\,}
\newcommand{\lsim}{\, \mbox{\raisebox{-1.ex}
{$\stackrel{\textstyle<}{\textstyle\sim}$}}\,}

\begin{document}
\title{\sc{Metric-Affine Gravity and Inflation}}

\author{Keigo \sc{Shimada}} 
 \email{kshimada@gravity.phys.waseda.ac.jp}
\affiliation{
Department of Physics, Waseda University,
Shinjuku, Tokyo 169-8555, Japan
}
\author{Katsuki \sc{Aoki}} 
 \email{katsuki-a12@gravity.phys.waseda.ac.jp}
\affiliation{
Department of Physics, Waseda University,
Shinjuku, Tokyo 169-8555, Japan
}
\author{Kei-ichi \sc{Maeda}} 
 \email{maeda@waseda.jp}
\affiliation{
Department of Physics, Waseda University,
Shinjuku, Tokyo 169-8555, Japan
}
\date{\today}
\preprint{WU-AP/1808/18}

\def\-#1{\frac{1}{#1}}

\def\eqm#1{\begin{eqnarray*} #1\end{eqnarray*}}
\def\eqn#1{\begin{eqnarray} #1\end{eqnarray}}
\def\d{\partial}
\def\l{\left}\def\r{\right}
\def\mat#1#2{\left(
	\begin{array}{#1} #2 \end{array}\right)}
\def\pic#1#2{　
	\begin{figure}[t]
	 \begin{center}
	   \includegraphics[width=120mm]{#1.eps}
	 \end{center}
	 \label{fig:}
	\captionsetup{labelformat=empty,labelsep=none}
	\caption{#2}
	\end{figure}}

\def\cf#1{\mathcal #1}

\def\subeqn#1{\begin{subequations}\begin{eqnarray}#1\end{eqnarray}\end{subequations}}
\def\seteq#1{\setcounter{equation}{#1}}
\def\nn{\nonumber}
\def\num#1{$#1$\hfill\refstepcounter{equation}(\theequation)}

\begin{abstract}
We classify the metric-affine theories of gravitation, in which the metric and 
the connections are treated as independent variables, by use of several constraints on the connections.
Assuming the Einstein-Hilbert action, 
we find that the equations for the distortion tensor (torsion and non-metricity) become algebraic, 
which means that those variables are not dynamical.
As a result, we can rewrite the basic equations in the form of Riemannian geometry.
Although all classified models recover the Einstein gravity in the Palatini formalism (in which we assume 
there is no coupling between matter and the connections), but   when matter field couples to the connections,
the effective Einstein equations include
an additional hyper energy-momentum tensor obtained from the distortion tensor.
Assuming a simple extension of a minimally coupled scalar field in metric-affine gravity, 
we analyze an inflationary scenario.  
Even if we adopt a chaotic inflation potential, 
certain parameters could satisfy observational constraints. 
 Furthermore, we find that a simple form of Galileon scalar field in metric-affine could cause G-inflation. 
\end{abstract}
\maketitle

\section{Introduction}

General relativity (GR) is undoubtedly one of the most successful relativistic gravitational theories since its proposal over a century ago. Countless experiments have been conducted to confirm its viability throughout the years (see for example \cite{Will:2014kxa,Will:2018bme}).  However recent 
 observations of the universe such as the acceleration of cosmic expansion 
 suggest  an alternative theory of gravity\cite{Riess:1998cb,Aghanim:2018eyx}. 
 In the early stage of the universe, we may also expect an accelerating expansion of the universe, the so-called 
 inflation\cite{Starobinsky:1980te,Guth:1980zm,Sato:1980yn,Hawking:1982cz,Guth:1982ec,
 Starobinsky:1982ee,Mukhanov:1982nu}(For many reviews of inflation see for example \cite{Martin:2013tda,Martin:2013nzq}).  
 Although we do not know the origin of the  inflaton field, which is responsible for the accelerating expansion 
 in the early stage of the universe, one possible answer could be modification of a gravitational theory
 such as Higgs inflation model \cite{spokoiny1984inflation, Maeda:1988rb, futamase1989chaotic,Bezrukov:2007ep,Bezrukov:2008ej,DeSimone:2008ei,
 Bezrukov:2009db,Bezrukov:2013fka,Germani:2010gm,Germani:2010gm,Germani:2010ux,Sato:2017qau}.
Although we have not yet had a satisfactory viable explanation to solve inflation as well as dark energy 
 in the framework of modern physics, we recently find 
 an astounding increase in the proposal of modified theories of gravity 
 \cite{Clifton:2011jh,Joyce:2014kja}.  By considering alternative theories of gravity, one may not only find a solution to these problems but also enforce our understanding of gravity itself.

 It is more than common to consider a purely metric theory with Riemannian geometry  when formulating alternative gravitational theories. This is more than natural since the best gravitational theory we know, the General Theory of Relativity, is written in terms of Riemannian geometry. However, one goes beyond Riemannian geometry and allow new structures to be included in a gravitational theory.  These theories constructed from non-Riemannian geometry may naturally exhibit new features into a theory.  Furthermore, one must note that some formalisms give an equivalent theory as GR, for example, teleparallel gravity \cite{Unzicker:2005in} or symmetric teleparallel gravity\cite{Nester:1998mp}. However, once we try to go to alternative theories of gravity, such as higher dimensions or non-minimal couplings, 
 these formalisms could differ from their purely metric counterparts, and provide some different solutions\cite{Sotiriou:2006qn, Exirifard:2007da, Dadhich:2010dg,Olmo:2011uz,Cai:2015emx,BeltranJimenez:2017tkd}. 
 
 In this paper, we will use a formalism called metric-affine geometry, in which the metric and the connection are independent variables \cite{Einstein:1923,Einstein:1925,Ferraris:1982am,Hehl:1994ue,Vitagliano:2010sr,Blagojevic:2012bc,Vitagliano:2013rna}. 
We consider theories given only by the curvatures, but not by the function of the 
 connections such as torsions. In particular, we mostly assume the Einstein-Hilbert action.

Keeping the above in mind, another interesting possibility to consider is a  scalar-tensor theory which 
has come to be popular throughout this decade. The idea behind these is relatively simple. To explain the unknown phenomena, e.g.,  inflation, dark matter, or dark energy, one could add an extra scalar degree of freedom to the two tensor degrees of freedom of gravity so that the problems are solved\cite{Brans:1961sx,Fujii:2003pa}.  However, most of these researches are done through the purely metric approach in which the geometry is Riemannian.  There are extensions to a non-Riemannian case \cite{Lindstrom:1976pq,Li:2012cc,Valdivia:2017sat,  Davydov:2017kxz,Aoki:2018lwx}, however, the fully metric-affine formulated theory and, more importantly, their applications to cosmology are yet to be explored. {\it This is the main purpose of this paper. }

Moreover, recently inflationary scenarios in Palatini/metric-affine theories have gained increasing attention, due to the fact that observational variables are different compared to the Riemannian case\cite{Tenkanen:2017jih,Markkanen:2017tun,Antoniadis:2018yfq,Antoniadis:2018ywb}. These seem to have other interesting features such as attractors\cite{Jarv:2017azx} and it's multifield extensions\cite{Carrilho:2018ffi}. By considering the inflaton as the Higgs, one could also consider the Palatini formalism of Higgs inflation\cite{Bauer:2008zj,Bauer:2010jg,Rasanen:2017ivk}. There are also approaches to consider inflation with a purely an affine approach \cite{Azri:2017uor,Azri:2018qux}.

The organization of this paper is as follows. We start with briefly classifying metric-affine theories of gravity by
some conditions on the connections  in $\S$II. 
 We then proceed to discuss a minimally coupled scalar field in the metric-affine formalism in
 $\S$III. The solutions for the connection of all classified models are given. 
Substituting the solution into the action, we find 
  the effective action described in the form of Riemannian geometry. Then we apply our formulation to two inflationary models in $\S$IV. One is a model with a 'minimally' coupled
scalar field, and another is G-inflation in which mimics the action introduced in \cite{Kobayashi:2010cm}. We show that observational constraint from the Planck 2018 results \cite{Aghanim:2018eyx} are satisfied 
by the appropriate choice of the coupling parameters. Summary and discussions are presented in $\S$V.  Some further extension of d'Alembertian  is also discussed in the Appendix.

\section{Classification of Metric-Affine Gravity}

\subsection{torsion, non-metricity and distortion tensor}
We shall start by classifying the metric-affine gravity theories, in which the (Riemann) metric and the affine connection are treated as independent variables. 
Since the precise definition of geometrical variables is crucial in metric-affine gravity theories, 
 this section is dedicated to clarifying how those variables are defined in metric-affine gravity
 and classify several approaches.

Once  a  connection $\Gamma$ is introduced,
 the covariant derivative is naturally defined by 
\beann
\Gnabla_\mu A_\nu &\equiv& \d_\mu A_\nu -{\Gamma^\alpha}_{\mu\nu}A_\alpha
\,,\\
\Gnabla_\mu B^\nu &\equiv& \d_\mu B^\nu +{\Gamma^\nu}_{\mu\alpha}B^\alpha
\,.
\enann

The most general connection in metric-affine gravity theories 
consists of the Levi-Civita connection as denoted by $\l\{\substack{\alpha\\\beta\gamma}\r\}$, which is fixed by the metric as 
\beann
\l\{\substack{\alpha\\ \mu\nu}\r\}\equiv 
{1\over 2}g^{\alpha\beta}\left(\partial_\mu g_{\beta\nu}
+\partial_\nu g_{\mu\beta}-\partial_\beta g_{\mu\nu}
\right)
\,,
\enann
 and 
 two additional geometrical tensors;  the torsion $
{{\cal T}^\lambda}_{\mu\nu}\equiv2{\Gamma^\lambda}_{[\mu\nu]}$
 and the 
 non-metricity  ${{\cal Q}_\alpha}^{\beta\gamma}\equiv \Gnabla_\alpha g^{\beta\gamma}$.
 Here the anti-symmetrization square brackets withholds a factor of $\-2$, i.e., 
 $[A,B]={1\over 2}(AB-BA)$.
Torsion ${{\cal T}^\lambda}_{\mu\nu}$ could normally be interpreted as a measure of how a vector is 'twisted' on a curved space when it is parallel transported, whereas non-metricity 
${{\cal Q}_\alpha}^{\beta\gamma}$ could be thought as for how the length of the vector changes through parallel transport.\\
\indent The Riemann curvature is defined by the connection  ${\Gamma^\alpha}_{\beta\gamma}$ as
\beann
{\GR}{}^\alpha_{~\beta\mu\nu}&\equiv&R^\alpha_{~\beta\mu\nu}(\Gamma)
\\
&=&\d_\mu {\Gamma^\alpha}_{\nu\beta}-\d_\nu {\Gamma^\alpha}_{\mu\beta}
+{\Gamma^\alpha}_{\mu\lambda}{\Gamma^\lambda}_{\nu\beta}-{\Gamma^\alpha}_{\nu\lambda}{\Gamma^\lambda}_{\mu\beta}\,.
\enann
Note that, by definition, it is only antisymmetric in the last two indices.\\
\indent 
We introduce  the distortion tensor $\kappa$ defined by
\beann
{\kappa^\alpha}_{\beta\gamma}\equiv{\Gamma^\alpha}_{\beta\gamma}-\l\{\substack{\alpha\\\beta\gamma}\r\}
\,,
\enann
which makes calculations in metric-affine gravity theories greatly simplified.
The distortion tensor is given by the torsion and non-metricity as
\beann
{\kappa^\lambda}_{\mu\nu}&=&\frac12\l({{\cal T}^\lambda}_{\mu\nu}+{{{\cal T}_\nu}^\lambda}_\mu-{{\cal T}_{\mu\nu}}^\lambda +{{{\cal Q}_\nu}^\lambda}_\mu+{{\cal Q}_{\mu\nu}}^\lambda-{{\cal Q}^\lambda}_{\mu\nu}\r)\,,
\enann
where the indices have been raised by the metric $g^{\mu\nu}$.

Inversely we have
\beann
{{\cal T}^\lambda}_{\mu\nu}&=&2{\kappa^\lambda}_{[\mu\nu]}
\\
{{\cal Q}_\alpha}^{\beta\gamma}&=&2\kappa^{(\beta\ \gamma)}_{\ \ \alpha}
\enann
By use of the distortion tensor $\kappa$, the Riemann tensor is decomposed as 
\begin{widetext}
\bea
{\GR}{}^\alpha_{~\beta\mu\nu}&=&{R^\alpha}_{\beta\mu\nu}+\nabla_\mu {\kappa^\alpha}_{\nu\beta}-\nabla_\nu {\kappa^\alpha}_{\mu\beta}+{\kappa^\alpha}_{\mu\lambda}{\kappa^\lambda}_{\nu\beta}-{\kappa^\alpha}_{\nu\lambda}{\kappa^\lambda}_{\mu\beta},
\label{eq:RiemKap}
\ena
\end{widetext}
where  ${R^\alpha}_{\beta\mu\nu}={R^\alpha}_{\beta\mu\nu}(\{\})$ is the Riemann tensor defined by the Levi-Civita connection $\{\}$, and 
$\nabla_\mu$ is the covariant derivative with respect to the Levi-Civita connection.
We also have two Bianchi identities:
\beann
{\GR{}^\lambda}_{[\alpha\beta\gamma]}&=&\nabla_{[\alpha}{{\cal T}^\lambda}_{\beta\gamma]}
-{{\cal T}^\sigma}_{[\alpha\beta}{{\cal T}^\lambda}_{\gamma]\sigma},\\
\Gnabla_{[\alpha}{\GR{}^\mu}_{|\nu|\beta\gamma]}&=&{{\cal T}^\lambda}_{[\alpha\beta}{\GR{}^{\mu}}_{|\nu|\gamma]\lambda}
\,.
\enann
Because of these identities, one must take extra care when deriving 
 energy-momentum  conservation   laws\cite{Hehl:1994ue}.\\

\subsection{Projective Invariance}

\indent In what follows, as for the curvature term, we consider the Einstein-Hilbert action
\eqn{
S_{\rm g}=\frac{M^2_{\text{Pl}}}2 \int \text{d}^4 x \sqrt{-g}\GR(g,\Gamma)\,,
\label{EH_action0}
}
where  $M^2_{\text{Pl}}\equiv\frac1{8\pi G}$  is the reduced Planck mass and $\GR(g,\Gamma)\equiv g^{\mu\nu}{\GR{}^\lambda}_{\mu\lambda\nu}$ is 
 the Ricci scalar.  Note that the Ricci scalar is uniquely defined
 by the contraction of the Riemann tensor.
 One must also keep in mind, however, that the Ricci tensor is not uniquely defined 
because   
the Riemann tensor does
not have  the usual (anti-) symmetries and 
the different contractions become possible.
It may compute different results. \\

\indent The action (\ref{EH_action0}) can be decomposed into the purely metric part and the distortion tensor as, 
\beann
S_{\rm g}&=&\frac{M^2_{\text{Pl}}}2 \int \text{d}^4 x \sqrt{-g}\Big[R(g)+g^{\mu\nu}\l({\kappa^\lambda}_{\mu\nu}{\kappa^\sigma}_{\sigma\lambda}-{\kappa^\lambda}_{\sigma\mu}{\kappa^\sigma}_{\nu\lambda}\r)\Big]\,,
\label{EH_action1}
\enann
 where we have dropped the surface terms appeared from the 2nd and 3rd terms in the rhs of Eq. (\ref{eq:RiemKap}). 
 This description shows that 
this model describes the Einstein gravity in the Riemannian geometry with 
the distortion tensor field $\kappa$.
In fact, by taking the variation of the action with respect to 
the metric $g^{\mu\nu}$ and the distortion tensor $\kappa$, we find
\bea
\frac{\delta S}{\delta g^{\mu\nu}}&=&\frac{M^2_{\text{Pl}}}2
\Big[G_{\mu\nu}+{\kappa^\lambda}_{\mu\nu}{\kappa^\sigma}_{\sigma\lambda}-{\kappa^\lambda}_{\sigma\mu}{\kappa^\sigma}_{\nu\lambda}
\nonumber 
\\
&&
-{1\over 2}g_{\mu\nu}
g^{\alpha\beta}\l({\kappa^\lambda}_{\alpha\beta}{\kappa^\sigma}_{\sigma\lambda}-{\kappa^\lambda}_{\sigma\alpha}{\kappa^\sigma}_{\beta\lambda}\r)
\Big]\,,
\label{variation_metric}
\ena
and
\bea
\frac{\delta S}{\delta {\kappa^\alpha}_{\beta\gamma}}&=&\frac{M^2_{\text{Pl}}}2 
\left[g^{\beta\gamma}{\kappa^\sigma}_{\sigma\alpha}+\delta^\beta_\alpha{\kappa^{\gamma\sigma}}_\sigma-{\kappa^{\beta\gamma}}_{\alpha}-\kappa^{\gamma\ \beta}_{\ \alpha}\right]\,.\nn\\
\label{variation_kappa1}
\ena
Note that there is no kinetic term of $\kappa$, which means 
it is not dynamical but fixed by the constraint equations. 
This characteristic will become crucial later on in the paper.

 \indent  In metric-affine geometry, there exists a new symmetry called 'projective symmetry' for the Einstein-Hilbert action (For geometrical aspects, see Sec. VI of the textbook \cite{RicciSchouten}).
The projective transformation is a transformation of the connection as
\eqn{
{\Gamma^\alpha}_{\beta\gamma}\to {{\tilde\Gamma}^\alpha}_{\beta\gamma}={\Gamma^\alpha}_{\beta\gamma}+\delta^\alpha_\gamma U_\beta
\,,
\label{projective_transformation}
}
which preserves the angle of two vectors and leaves the geodesic equation equivalent up to the redefinition of the affine parameter such that $\lambda\rightarrow \tilde \lambda(\lambda)$, 
where 
$\tilde \lambda$ is given by the solution of the differential  equation
\beann
{d^2\tilde \lambda\over d\lambda^2}-U_\beta{dx^\beta\over d\lambda}{d\tilde \lambda\over d\lambda}=0\,,
\enann
which could be integrated as below.
\eqm{
\tilde \lambda =\int e^{\int U_\beta dx^\beta }d\lambda \,.
}
Under this transformation, the Ricci scalar is invariant as 
\eqn{
\GR(g,\Gamma)\to\stackrel{\,\tilde\sGamma\,}{R}(g,\tilde\Gamma)=\GR(g,\Gamma)\,.
}
Thus the Einstein-Hilbert action has ``projective invariance''.
Note that for some gravitational action such as the curvature-squared gravity theory, 
such symmetry may no longer exist.

If the matter action also has the projective invariance, the full theory has such symmetry.
We call it a projective invariant theory.
Although the constraint equations for $\kappa$  do not fix all components of the connection,
this ambiguity does not appear in the basic equations.
It could be considered as a kind of gauge freedom.

We can see this fact explicitly as follows:
The variation with respect to the distortion tensor  provides 
the constraint equations for $\kappa$.
For simplicity, we consider the vacuum case or the model with 
matter field which does not couple to the connection (the so-called Palatini gravity theory).
 The variation with respect to the  distortion tensor leads to
\beann
g^{\beta\gamma}{\kappa^\sigma}_{\sigma\alpha}+\delta^\beta_\alpha{\kappa^{\gamma\sigma}}_\sigma-{\kappa^{\beta\gamma}}_{\alpha}-\kappa^{\gamma\ \beta}_{\ \alpha}=0\,,
\enann
which is solved as
\eqn{
{\kappa^\alpha}_{\beta\gamma}=\-4 \delta^\alpha_{~\gamma}\, \kappa_\beta
\,,
\label{trace_kappa}
}
where $ \kappa_\beta :=\kappa^\lambda_{~\beta\lambda}$.
This indicates that the distortion tensor $\kappa$, and thus the connection $\Gamma$, is undetermined up to
the trace $\kappa_\beta$.
These remaining four degrees of freedom correspond to the projective transformation vector $U_\beta$.
This result is also easily understood as follows 
 when we introduce the reduced distortion tensor defined by
\beann
{\bar \kappa^\alpha}_{~\beta\gamma}\equiv 
\kappa^\alpha_{~\beta\gamma}-{1\over 4} \delta^\alpha_{~\gamma}\, \kappa_\beta
\,,
\enann
which is, by definition, a trace-free tensor.\\
\indent Now, by the use of ${\bar \kappa^\alpha}_{~\beta\gamma}$, the action $S$ is rewritten as
\bea
S_{\rm g}&=&\frac{M^2_{\text{Pl}}}2 \int \text{d}^4 x \sqrt{-g}\Big[R(g)
\nn
\\
&&
+g^{\mu\nu}\l({\bar \kappa^\lambda}_{~\mu\nu}{\bar \kappa^\sigma}_{~\sigma\lambda}-{\bar \kappa^\lambda}_{~\sigma\mu} {\bar\kappa^\sigma}_{~\nu\lambda}\r)\Big]\,,
\label{EH_action2}
\ena
which gives the Einstein equations as 
\beann
G_{\mu\nu}={M^{-2}_{\text{Pl}}}\left[T_{\mu\nu}+\tau_{\mu\nu}\right]
\,,
\enann
where the energy-momentum tensor of usual matter field $T_{\mu\nu}$ is given by
\beann
T_{\mu\nu}=-2{\delta S_{\rm m}\over \delta g^{\mu\nu}}
\,,
\enann 
and the  hyper energy-momentum tensor $\tau_{\mu\nu}$ is defined by 
\beann
\tau_{\mu\nu}&\equiv&
-{\bar \kappa^\lambda}_{~\mu\nu}{\bar \kappa^\sigma}_{~\sigma\lambda}
+{\bar \kappa^\lambda}_{~\sigma\mu}{\bar \kappa^\sigma}_{~\nu\lambda}
\nonumber 
\\
&&
+{1\over 2}g_{\mu\nu}
g^{\alpha\beta}\l({\bar \kappa^\lambda}_{~\alpha\beta}{\bar \kappa^\sigma}_{~\sigma\lambda}-{\bar \kappa^\lambda}_{~\sigma\alpha}{\bar \kappa^\sigma}_{~\beta\lambda}\r)\,,
\enann
which can be treated as an additional energy-momentum tensor in the Riemannian geometry 
 coming from 
${\bar \kappa^\alpha}_{~\beta\gamma}$.

The variation with respect to $\kappa^\lambda_{~\mu\nu}$ is also given only by 
 $\bar \kappa^\lambda_{~\mu\nu}$ as
\bea
\frac{\delta S}{\delta \kappa^\alpha_{~\beta\gamma}}&=&\frac{M^2_{\text{Pl}}}2 
\left[g^{\beta\gamma}{\bar \kappa^\sigma}_{~\sigma\alpha}+\delta^\beta_\alpha{\bar \kappa^{\gamma\sigma}}_{~~\sigma}-{\bar \kappa^{\beta\gamma}}_{~~\alpha}-\bar \kappa^{\gamma\ \beta}_{\ \alpha}\right]\,.\nn\\
\label{variation_kappa2}
\ena

Hence the constraint equations are 
\bea
g^{\beta\gamma}{\bar \kappa^\sigma}_{~\sigma\alpha}+\delta^\beta_\alpha{\bar \kappa^{\gamma\sigma}}_{~~\sigma}-{\bar \kappa^{\beta\gamma}}_{~~\alpha}-\bar \kappa^{\gamma\ \beta}_{\ \alpha}=-2M^{-2}_{\text{Pl}}\frac{\delta S_{\rm m}}{\delta {\bar \kappa^\alpha}_{~\beta\gamma}}\,,\nn\\
\label{constraint_1}
\ena
and 
\bea
\frac{\delta S_{\rm m}}{\delta {\kappa}_{\beta}}=0\,.
\label{constraint_2}
\ena

When the matter action is projective invariant, 
Eq. (\ref{constraint_2}) becomes trivial.
If there is no coupling between matter field and the connection, the r.h.s. in Eq. (\ref{constraint_1})
vanishes. As a result, we obtain $\bar \kappa^\lambda_{~\mu\nu}=0$, and then find the conventional Einstein gravity theory with the Levi-Civita connection.
When the theory has matter field coupled with the distortion tensor
 $ \kappa^\lambda_{~\mu\nu}$,
we have an additional contribution of  $\bar \kappa^\lambda_{~\mu\nu}$, which is determined 
by the constraint equation (\ref{constraint_1}).
We classify this projective invariant gravity theory as Model I.
\\~

\subsection{Non-projective invariant gravity theories}

Since matter, in general, is not projective invariant, all components of the connection
 should be fixed.  
Hence when we discuss non-projective invariant gravity theories,
we may impose an additional constraint on the connections 
to eliminate the unfixed components in the Einstein-Hilbert action. (For further of consequences of fixing the projective gauge, see for example \cite{Iosifidis:2018jwu})
There are the following two common approaches. 
One is to take the torsion-free  condition (${{\cal T}^\lambda}_{\mu\nu}=0$), and 
the other is to take the metric-compatible condition
 (${{\cal Q}_\alpha}^{\beta\gamma}=0$). In general, these two conditions do not have to simultaneously coincide in general.
 We classify  these cases into  Model II(a) and Model II(b), respectively. 
Both of these conditions are commonly assumed {\it a priori} in the gravitational  action. 
As we will see below, for the Einstein-Hilbert action, both approaches compute the Einstein equations and the Levi-Civita connection if matter does not couple to the connection.  

For  Model II(a), noting that the distortion is restricted as 
${\kappa^\lambda}_{[\mu\nu]}={{\cal T}^\lambda}_{\mu\nu}/2=0$,
 the solution (\ref{trace_kappa}) for the constraint equation for the connection reads
\eqn{
{\kappa^\alpha}_{\beta\gamma}=0 
\label{eq:LCans}\,,
}
which gives the Levi-Civita connection.

Similarly, for  Model II(b), since $\kappa^{(\beta\ \gamma)}_{\ \ \alpha}={{\cal Q}_\alpha}^{\beta\gamma}/2=0$,
the constraint equations in the metric-compatible case reads
Eq. (\ref{eq:LCans}).
We again find the Levi-Civita connection.

When the matter couples with the connection, 
the reduced distortion tensor $\bar \kappa$  and 
the trace $\kappa_\mu$ are obtained by the constraint equations
 (\ref{constraint_1}) and  (\ref{constraint_2}).
The hyper energy-momentum tensor appears in the Einstein equations.
Although the trace term of the distortion tensor is also fixed, it does not appear
in the Einstein equations.

In Appendix, we present the explicit description of the Einstein equations and 
the constraint equations by use of the non-metricity tensor ${{\cal Q}_\alpha}^{\beta\gamma}$
 for Model II(a) 
and the torsion tensor  ${{\cal T}_\alpha}^{\beta\gamma}$  for Model II(b), respectively.

Although the above ansatz of the torsion-free or the metric compatibility 
provides a consistent gravity theory, when we break the projective invariance, 
such a condition may be too strict because we have only
 four undetermined components in the connection. 
The minimum condition that one could impose is constraining some vector ${\cal C}^\mu$, which consists of the distortion tensor, via a Lagrange multiplier, as
\begin{widetext}
\begin{eqnarray}
S_{\rm g}=\frac{M^2_{\text{Pl}}}2 \int \text{d}^4 x \sqrt{-g}\l\{R(g)+g^{\mu\nu}\l({\kappa^\lambda}_{\mu\nu}{\kappa^\sigma}_{\sigma\lambda}-{\kappa^\lambda}_{\sigma\mu}{\kappa^\sigma}_{\nu\lambda}\r)+\lambda^\mu {\cal C}_\mu(\kappa)\r\}\,.
\label{EH_action_LM}
\end{eqnarray}
\end{widetext}
The variation of the Lagrange multiplier $\lambda^\mu$ gives four constraint equations
 ${\cal C}_\mu(\kappa)=0$,
which fixes four undetermined components in the connection.

When we perform a projective transformation (\ref{projective_transformation}), 
we find 
\beann
{\cal T}^\lambda_{~\mu\nu}&\rightarrow& \tilde{\cal T}^\lambda_{~\mu\nu}
= {\cal T}^\lambda_{~\mu\nu}+\Delta {\cal T}^\lambda_{~\mu\nu}= {\cal T}^\lambda_{~\mu\nu}+2\delta^\lambda_{[\nu}U_{\mu]}\\
{\cal Q}^\lambda_{~\mu\nu}&\rightarrow&
\tilde  {\cal Q}^\lambda_{~\mu\nu}=
 {\cal Q}^\lambda_{~\mu\nu}+
 \Delta {\cal Q}^\lambda_{~\mu\nu}=
 {\cal Q}^\lambda_{~\mu\nu}+2g_{\mu\nu}U^\lambda
\,,
\enann
which give  $U_\mu=\Delta {\cal T}^\lambda_{~\mu\lambda}/3=\Delta {\cal Q}^\lambda_{~\mu\lambda}/2=\Delta {\cal Q}_{\mu~\lambda}^{~\lambda}/8$.
Hence in order to break the projective invariance, we could choose
the constrained vector ${\cal C}_\mu(\kappa)$ either of the following three vectors; 
\\
(a) the torsion vector: ${\cal T}_\mu\equiv {{\cal T}^\lambda}_{\mu\lambda}$, \\
(b) the non-metricity trace vector : $ {\cal Q}_\mu\equiv {\cal Q}^\lambda_{~\mu\lambda}$ , 
\\
(c) the Weyl vector : ${\cal W}_\mu\equiv\-4  {\cal Q}_{\mu~\lambda}^{~\lambda}$. \\
We will classify these models III (a), III (b) and III (c), respectively.
The constraint equation ${\cal C}_\mu(\kappa)=0$ in each case gives
 $\kappa^\beta_{~\alpha\beta}=0$
because ${\cal T}_\mu=2{\kappa^\lambda}_{[\mu\lambda]}$, 
$ {\cal Q}^\mu=2\kappa^{(\lambda\ \mu)}_{\ \ \lambda}$,
and  ${\cal W}_\mu=\frac12\kappa^\lambda_{~\mu\lambda}$. As a result, if matter does not couple with 
the connection, we find the conventional Einstein equations with the Levi-Civita connection.
In the case with matter coupled the connection, we have to include the modified hyper energy-momentum tensor 
in the basic equations. The modification of the hyper energy-momentum tensor appears because of the Lagrange multiplier term.
When we take the variation of the action $S$ with respect to the metric, we find 
the Einstein equations as
\beann
&&
\\
&&G_{\mu\nu}=M_{\text{Pl}}^{-2}\left(T_{\mu\nu}+
\tau_{\mu\nu}+\Delta\tau_{\mu\nu}\right)
\,,
\enann
where 
\beann
\Delta\tau_{\mu\nu}=-\lambda^\alpha{\delta C_\alpha\over \delta g^{\mu\nu}}
\,.
\enann
For  Models III(a) and (c), $C_\alpha$ does not contain the metric $g$, then 
the modification term vanishes. While for  Model III (b), since 
$ {\cal Q}_\lambda=g_{\alpha\beta}g^{\lambda\sigma}\kappa^{\beta}_{~\lambda\sigma}+\kappa^{\lambda}_{~\mu\lambda}$,  
$\Delta\tau_{\mu\nu}$ is not trivial.
We find
\beann
\Delta\tau_{\mu\nu}=-\lambda_{(\mu}\kappa_{\nu)~\lambda}^{~\lambda}+\lambda_\alpha
\kappa^\alpha_{~(\mu\nu)}
\,.
\enann
The constraint equation for $\kappa$ is 
\bea
&&g^{\beta\gamma}\kappa^\sigma_{~\sigma\alpha}
+\delta^\beta_\alpha\kappa^{\gamma\sigma}_{~~\sigma}
-\kappa^{\beta\gamma}_{~~\alpha}- \kappa^{\gamma\ \beta}_{\ \alpha}
\nonumber \\
&&~~~~~~~~~~~~+\lambda^\mu{\delta C_{\mu}\over \delta \kappa^\alpha_{~\beta\gamma}}
=-2M^{-2}_{\text{Pl}}\frac{\delta S_{\rm m}}{\delta \kappa^\alpha_{~\beta\gamma}}\,,
\label{constraint_L}
\ena
where
\beann
\lambda^\mu{\delta C_{\mu}\over \delta \kappa^\alpha_{~\beta\gamma}}=\left\{
\begin{array}{cc}
\delta_\alpha^{~\gamma}\lambda^\beta-\delta_\alpha^{~\beta}\lambda^\gamma&({\rm Model~ III(a)})  \\[.5em]
\delta_\alpha^{~\beta}\lambda^\gamma+ g^{\beta\gamma}\lambda_\alpha&({\rm Model~ III(b)})   \\[.5em]
{1\over 2}\delta_\alpha^{~\gamma}\lambda^\beta &({\rm Model~ III(c)})  \\
\end{array} 
\right.
\,.
\enann

Solving the constraint equation (\ref{constraint_L}) with ${\cal C}_\mu(\kappa)=0$,
we obtain the distortion tensor $\kappa^\lambda_{~\mu\nu}$ and the Lagrange multiplier $\lambda^\mu$,
which fix the hyper energy-momentum tensor ( and its modification).

If there is no coupling between matter field and the connection, 
since $\kappa^\lambda_{~\mu\nu}=0$ and $\lambda^\mu=0$,
we again recover the conventional Einstein equations with Levi-Civita connection.

\begin{widetext}

We summarize the classification of metric-affine gravity theories in Table \ref{classification}.

\begin{table}[h]
\begin{center}
\begin{tabular}{c|c|c|c|c}
\hline
Models&constraint&properties&Palatini formalism&metric-affine formalism\\
\hline\hline
&&&&\\[-1em]
I &\large {${\delta S_G}/{\delta \Gamma^\lambda_{~\mu\nu}}\,\delta^\lambda_\nu=0$}&projective invariant&&\\
\cline{1-3}
&&&&\\[-1em]
II (a)&${{\cal T}^\lambda}_{\mu\nu}=0$&torsion-free&&\\
II (b) &$ {{\cal Q}^\lambda}_{\mu\nu}=0$&metric compatible&Einstein equations&$+\tau_{\mu\nu}$\\[.1em] 
\cline{1-3}
&&&$G_{\mu\nu}(g)=M_{\rm Pl}^{-2}T_{\mu\nu}$&\\[-1em]
III (a) &${\cal T}^\lambda=0$&&&\\
\cline{5-5}
III (b) &$ {\cal Q}^\lambda=0$&${{\cal T}^\lambda}_{\mu\nu}\neq 0$ 
and $ {{\cal Q}^\lambda}_{\mu\nu}\neq 0$&&$+\tau_{\mu\nu}+\Delta \tau_{\mu\nu}$\\
\cline{5-5}
III (c) &${\cal W}^\lambda=0$&{\rm in ~general}
&&$+\tau_{\mu\nu}$
\\
\hline
\end{tabular}
\caption{The classification of metric-affine gravity theories. For the Einstein-Hilbert action, the basic equations turn out to be just
the Einstein equations if the connection does not coupled to matter field (Palatini formalism), 
but there appears an additional term (the hyper energy-momentum tensor) from the distortion tensors in general metric-affine gravity theories.}
\label{classification}
\end{center}
\end{table}

\end{widetext}

\section{Scalar Field in Metric-Affine Gravity}
\label{scalar_field_metric_affine}
\subsection{``canonical'' scalar field}
Here we consider a minimally coupled ``canonical'' scalar field in
 the metric-affine formalism.
 The action for a real scalar field in a flat Minkowski space is given by
 \eqn{
S_{\phi,\text{flat}}=\int d^4x \l(-\-2 \eta^{\mu\nu}\d_\mu \phi \d_\nu\phi -V(\phi)\r)
\,,
\label{flat_action_1}
}
which can be rewritten by integration by parts to the equivalent action, up to the surface term, as
\eqn{
S_{\phi,\text{flat}}=\int d^4x \l(\-2\phi \fsquare\phi-V(\phi)\r)\,,
\label{flat_action_2}
}
where $ \fsquare\equiv \eta^{\mu\nu}\partial_\mu\partial _\nu$ is	 the flat-space d'Alembertian operator. 

When we discuss a scalar field in a curved spacetime, we have to extend the above action in a covariant form. In Riemannian geometry, covariantization is straightforward. One simply has to substitute the volume density $\sqrt{-g}$ and replace $\partial_\mu$ with $\nabla_\mu$.
 The result of the covariantization of  (\ref{flat_action_1}) is equivalent to the covariantization of  (\ref{flat_action_2}) up to the surface term. However, in metric-affine geometry, as we will see, not only two covariantizations give the different results, but also the covariantization of (\ref{flat_action_2}) is not trivial.
\\
\indent If we start from the action (\ref{flat_action_1}), 
the scalar field does not couple to the connection since 
 $\Gnabla_\mu\hskip -.3em\phi=\d_\mu\phi$. 
On the other hand, when we covariantize the action (\ref{flat_action_2}), 
there exists some ambiguity in the definition of the d'Alembertian operator $\Gsquare$
in a metric-affine curved spacetime. In the presence of non-metricity, one can construct two different second-order covariant derivative operators; $\Gnabla_\mu\Gnabla\hskip -.3em{}^\mu$ and $\Gnabla\hskip -.3em{}^\mu\hskip -.4em\Gnabla_\mu$
 with $\Gnabla\hskip -.3em{}^\mu:=g^{\mu\nu}\hskip -.3em\Gnabla_\nu$. 
 As it will be shown the two actions (\ref{flat_action_1}) and (\ref{flat_action_1}), even though equivalent in flat space-time, differ greatly in metric-affine curved space.\\
  By imposing $\Gsquare \to \fsquare$ in the limit of a flat spacetime, the d'Alembertian operator in a curved spacetime could be defined as 
\bea
\Gsquare &=&\alpha\Gnabla\hskip -.3em{}^\mu\hskip -.3em\Gnabla_\mu+(1-\alpha)\Gnabla_\mu\Gnabla\hskip -.3em{}^\mu
\,,
\label{Gsquare}
\ena
where $\alpha$ labels the difference between the two operators. \\

Thus we suggest that the action for a scalar field  in a curved spacetime is described as
\eqn{
S_{\phi}=\int d^4x\sqrt{-g} \l(\-2\phi\Gsquare\phi-V(\phi)\r)
\,,
\label{eq:action_phi}
}which is the covariant version of (\ref{flat_action_2}).\\
Since  $\Gsquare \phi$ is described as
\bea
\Gsquare\phi &=&\square\phi +\left[(1-\alpha)g^{\alpha\beta}\kappa^\gamma_{~\gamma\beta}
-\alpha g^{\beta\gamma}\kappa^\alpha_{~\beta\gamma}\right]\partial_\alpha\phi
\nonumber \\
&=&
\square\phi -\left(\alpha{\cal Q}^\lambda-2{\cal W}^\lambda+{\cal T}^\lambda\right)\partial_\lambda \phi
\,,
\label{Gsquare_alpha}
\ena
where 
$\square\phi:={\displaystyle {1\over \sqrt{-g}}\partial_\mu  (\sqrt{-g}g^{\mu\nu}\partial_\nu \phi )}$, 
the variation  with respect to the distortion tensor gives
\eqn{
\frac{\delta S_{\phi}}{\delta {\kappa^\alpha}_{\beta\gamma}}&=&-{\alpha\over 2}g^{\beta\gamma}\phi \d_\alpha \phi +{(1-\alpha)\over 2}
\delta^\beta_\alpha \phi \d^\gamma\phi
\,.
\label{Sphidelkappa}
}

Now we solve the constraint equation for $\kappa^\lambda_{~\mu\nu}$ in each model.
\subsubsection{\rm Model I}
First one must note that, in general, metric-affine gravitational theories are not manifestly projectively invariant. So in order to constrain the theory to 'become' projective invariant, one must impose certain conditions as will be shown below.\\
\indent When we perform the projective transformation (\ref{projective_transformation}), 
we find 
\beann
\Gsquare\phi\rightarrow \Gsquare\phi+(1-2\alpha)U^\lambda\partial_\lambda \phi
\,.
\enann
Thus by fixing $\alpha$ to the value of $\alpha=1/2$, the theory becomes projective invariant.\\
\indent Now by solving the constraint equation 
\eqn{
&&\frac{M^2_{\text{Pl}}}2\l(g^{\beta\gamma}{\bar \kappa^\sigma}_{~\sigma\alpha}+\delta^\beta_\alpha{\bar \kappa^{~\gamma\sigma}}_\sigma-{\bar \kappa^{~\beta\gamma}}_{\alpha}-\bar \kappa^{\gamma\ \beta}_{\ \alpha}\r)\nn\\
&&\qquad\qquad\qquad-{1\over 4}g^{\beta\gamma}\phi \d_\alpha \phi +{1\over 4}
\delta^\beta_\alpha \phi \d^\gamma
\phi=0
\,,
}
the connection is obtained as
\eqn{
{\bar \kappa^\alpha}_{~\beta\gamma}=\frac{\phi}{4M_{\text{Pl}}^2}\l(\delta^\alpha_\beta\d_\gamma\phi-g_{\beta\gamma}\d^\alpha\phi\r)\,,\label{eq:PI1}
}
The torsion and non-metricity are given by 
\eqn{
{{\cal T}^\lambda}_{\mu\nu}&=&\frac{\phi}{2M_{\text{Pl}}^2}\delta^\lambda_{~[\mu}\d_{\nu]}\phi,\label{eq:PI2}\\
{{\cal Q}_\lambda}^{\mu\nu}&=&0,
}up to gauge freedom. This shows that for a projective invariant 'minimally' coupled scalar
field, there is a gauge that allows the connection to be metric-compatible while there is none that cancels out torsion. This is different from metric-affine $f(R)$ theory which is similarly projective invariant but admits both a metric-compatible gauge and a torsion-free gauge \cite{Sotiriou:2006qn}. Also, note that under projective transformation one could also obtain Weyl geometry since non-metricity changes as
\beann
{\cal Q}_\alpha^{~\beta\gamma}\overset{\Gamma\to\tilde\Gamma}\to
\tilde {\cal Q}_\alpha^{~\beta\gamma}=2U_\alpha g^{\beta\gamma}=8{\cal W}_\alpha g^{\beta\gamma}\,.
\enann
For history and recent progress in Weyl geometry see for example \cite{Scholz:2011za}

Now we find that the Euler-Lagrangian equation of the distortion tensor is algebraic and 
then it does not introduce new degrees of freedom, 
whereas the equations for the metric and scalar field carry the degrees of freedom. 
In such a case, inserting the solution of  the distortion tensor into the actions (\ref{EH_action2}) and (\ref{eq:action_phi}),
one could obtain an effective Lagrangian.
As a result, we find the total action $S_{{\rm g} \phi}:=S_{\rm g}+S_\phi$ purely in terms of the metric and the scalar field as
\eqn{
S_{{\rm g} \phi}&=&\int d^4x\sqrt{-g} \l[\frac{M_{\text{Pl}}^2}2R(g)\nn\r.\\
&&\l.\qquad-{1\over 2}\l(1-\frac{3\phi^2}{8M^2_{\text{Pl}}}\r)(\nabla\phi)^2-V(\phi)\r]
\,.\nn\\
\label{eq:PI3}
}
 Hence this model can be analyzed as a gravity theory with a scalar field that has a modified kinetic term in the usual Riemannian geometry formalism.\\
\subsubsection{\rm Models II (a) and (b)}
Now we consider the Einstein-Hilbert action and the same action of the scalar field (\ref{eq:action_phi}), 
but without imposing projective symmetry. In Model II(a), we instead impose the torsion-free condition ${\cal T}^\lambda_{~\mu\nu}=0$.  
The parameter $\alpha$ is not fixed because we do not assume projective invariance.

The constraint equation for the connection $\kappa^\lambda_{~\mu\nu}$ is now
\eqn{
&&\frac{M^2_{\text{Pl}}}2\l[g^{\beta\gamma}{\kappa^\sigma}_{\sigma\alpha}+\delta^{( \beta}_\alpha
\kappa^{\gamma ) \sigma}_{~~\sigma}-{\kappa^{(\beta\gamma)}}_{\alpha}-\kappa^{(\gamma\ \beta)}_{\ \alpha}\r]\nn\\
&&\qquad\qquad-{\alpha\over 2}g^{\beta\gamma}\phi \d_\alpha \phi +{(1-\alpha)\over 2}
\delta^{(\beta}_\alpha \phi \d^{\gamma)}
\phi=0
\,.\nn\\
}

Solving this constraint equation, we find the solution for the distortion as
\beann
\kappa^\alpha_{~\beta\gamma}&=&\frac{1}{6M^2_{\text{Pl}}}\l[3(\alpha-1)g_{\beta\gamma}\phi\d^\alpha\phi +2(\alpha+1)\delta^\alpha_{(\beta}\phi\d_{\gamma)}\phi\r]
\,,
\enann
which gives the non-metricity as
\beann
{\cal Q}_\lambda^{~\mu\nu}&=&\frac{1}{3M^2_{\text{Pl}}}\l[(\alpha+1)g^{\mu\nu}\phi\d_\lambda\phi +2(2\alpha-1)\delta_\lambda^{(\mu}\phi\d^{\nu)}\phi\r]
\,.
\enann
Inserting this solution into the original action, we again obtain an effective action written purely in Riemannian geometry, as
\bea
S_{{\rm g}\phi}&=&\int d^4x\sqrt{-g} \Big{[}\frac{M_{\text{Pl}}^2}2R(g)
\nonumber \\
&&
-{1\over 2}f(\phi)(\nabla\phi)^2-V(\phi)\Big{]}
\,,
\label{effective_action_IIa}
\ena
with 
\beann
f(\phi):= 1+\frac{(11\alpha^2-8\alpha-1)}{6M^2_{\text{Pl}}} \phi^2
\,.
\enann

  As for Model II(b), we assume ${\cal Q}^\lambda_{~\mu\nu}=0$, i.e., the metric-compatible condition is satisfied a priori. 
The solutions for the distortion  is given by
\beann
{\kappa^\alpha}_{\beta\gamma}=\frac{\phi}{4M_{\text{Pl}}^2}\l(\delta^\alpha_\beta\d_\gamma\phi-g_{\beta\gamma}\d^\alpha\phi\r)
\,,
\enann
which fixes the torsion as 
\beann
{{\cal T}^\lambda}_{\mu\nu}&=&\frac{\phi}{2M_{\text{Pl}}^2}\delta^\lambda_{~[\mu}\d_{\nu]}\phi\,.
\enann

One may first notice that the distortion of Model II(b) is the same form as the one from Model I (\ref{eq:PI1}). However, the latter admits gauge transformations, while the former does not.\\
\indent Furthermore, the equivalent action described in Riemannian geometry  becomes precisely the same as the previous one (\ref{eq:PI3}). 
\subsubsection{\rm Models III (a), (b) and (c)}
Now just as in the previous section we will consider constraining the geometry through Lagrange multipliers.\\
\indent Taking the variation of the Einstein-Hilbert action with the Lagrange multiplier (\ref{EH_action_LM}) plus the scalar field action (\ref{eq:action_phi}),
we find the constraint equation
\bea
&&g^{\beta\gamma}\kappa^\sigma_{~\sigma\alpha}
+\delta^\beta_\alpha\kappa^{\gamma\sigma}_{~~\sigma}
-\kappa^{\beta\gamma}_{~~\alpha}- \kappa^{\gamma\ \beta}_{\ \alpha}
+\lambda^\mu{\delta C_{\mu}\over \delta \kappa^\alpha_{~\beta\gamma}}
\nonumber \\
&& 
+M^{-2}_{\text{Pl}}\Big[
-\alpha g^{\beta\gamma}\phi \d_\alpha \phi +(1-\alpha)
\delta^{\beta}_\alpha \phi \d^{\gamma}
\phi\Big]=0
\,.
~~~~
\label{constraint_L2}
\ena
Contracting the above equation by $\delta_\alpha^{\gamma}$,
we find the Lagrange multiplier as
\beann
\lambda_\mu=\left\{
\begin{array}{cc}
{(2\alpha-1)\over 3M^{2}_{\text{Pl}}}\phi\partial_\mu \phi &({\rm Model~ III(a)})  \\[.5em]
{(2\alpha-1)\over 2M^{2}_{\text{Pl}}}\phi\partial_\mu \phi &(\text{Models III(b) and (c)})   
 \\
\end{array} 
\right.
\,.
\enann

Interestingly, we find that the results in Models III(a) and (b) are exactly the same as those in Models II(a) and (b), respectively.
We find the same connections $\kappa^\lambda_{~\mu\nu}$, and the 
equivalent action (\ref{effective_action_IIa}) in Riemannian geometry.\\
\indent On the other hand, as for Model III(c),
neither metric-compatibility nor the torsion-free condition are satisfied. 
The connection becomes
\bea
\kappa^\alpha_{~\beta\gamma}&=&\-{16M^2_{\text{Pl}}}\Big[2(2\alpha-3)g_{\beta\gamma}\phi\d^\alpha\phi 
\nn
\\
&+&(2\alpha+3)\delta^\alpha_{(\beta}\phi\d_{\gamma)}\phi +(6\alpha+1)\delta^\alpha_{[\beta}\phi\d_{\gamma]}\phi\Big]
\,,~~~
\ena
which gives the torsion and non-metricity as
\eqn{
{\cal T}^\alpha_{~\beta\gamma}&=&\frac{6\alpha+1}{8M_{\text{Pl}}^2}\delta^\alpha_{[\beta}\phi\d_{\gamma]}\phi,\\
{\cal Q}_\alpha^{~\beta\gamma}&=&\frac{2\alpha-1}{8M_{\text{Pl}}^2}\l(-g^{\beta\gamma}\phi\d_\alpha\phi+4\delta_\alpha^{(\beta}\phi\d^{\gamma)}\phi\r)
\,.
}
The resulting equivalent  action in Riemannian geometry is given by (\ref{effective_action_IIa}) 
with
\eqn{
f(\phi)&=&1+\frac{3(12\alpha^2-12\alpha-1)}{32M^2_{\text{Pl}}}\phi^2
}

\section{Applications to Inflation}

\subsection{chaotic inflation in metric-affine gravity}
\label{chaotic_inflation}
\indent \indent In the previous section we have shown that the metric-affine gravity theory
with a ``canonical'' scalar field
could be re-written to an equivalent Riemann geometrical action
(\ref{effective_action_IIa}) with
\eqn{
f(\phi)&=&1+B(\alpha)\frac{\phi^2}{M^2_{\rm Pl}}
\label{rel_fB}
\,,
}
where $B(\alpha)$ is given by
\eqn{
B(\alpha)=\begin{cases}
\frac16(11\alpha^2-8\alpha-1) &\text{Models II(a) and III(a)} \\
\qquad -\frac38 &\text{Models I, II(b), and III(b)}\\
\frac{3}{32}(12\alpha^2-12\alpha-1)&\text{Model III(c)}
  \end{cases}\nn
}
Note that $B(\alpha)\geq -\frac9{22} $ for Models II(a) and III(a) and  $B(\alpha)\geq -\frac38$
for Model III(c).
The function $B(\alpha)$ coincides at a single point for all models at $\alpha=1/2$ with $\large B\l(\frac12\r)=-\frac38$.
This function $B(\alpha)$ solely depends on what geometry is chosen within the framework of metric-affine geometry. For the rest of the paper $B(\alpha)$ would be re-taken as the parameter of the theories. \\
\indent The scalar field in the action (\ref{effective_action_IIa}) is canonically normalized 
by  the redefinition of the scalar  field as
\eqn{
\text{d}\Phi=\sqrt{1+B(\alpha)\frac{\phi^2}{M^2_{\rm Pl}}}\text{d}\phi
\,,
}
which could be integrated as
\beann
\Phi=\left\{
\begin{array}{lc}
{1\over 2 }\left[\phi\sqrt{1+{B\phi^2\over  M^2_{\rm Pl}}}+{M_{\rm Pl}\over  B^{1/2}}\sinh^{-1}\left({B^{1/2}\phi\over  M_{\rm Pl}}\right)\right]  &(B>0)  \\[1em]
~~~~~~~~~~~~~~~~~~~~~~~~~\phi& (B=0)\\[1em]
 {1\over 2 }\left[\phi\sqrt{1-{|B|\phi^2\over  M^2_{\rm Pl}}}+{M_{\rm Pl}\over  |B|^{1/2}}\sin^{-1}\left({|B|^{1/2}\phi\over  M_{\rm Pl}}\right)\right]   &(B<0)   \\
\end{array} 
\right.
\,.
\enann
Thus, instead of a modified kinetic term, the action is simply described by a canonical single field $\Phi$ with, as a consequence, a deformed potential in disguise,
\eqn{
S_{{\rm g}\Phi}&=&
\int \text{d}^4x\sqrt{-g}\Big[\frac{M^2_{\text{Pl}}}2R(g)
\nn
\\
&&
~~~~~~~~~~-\frac12 \d_\mu \Phi\d^\mu\Phi -U(\Phi))\Big]
\,,
\label{action_inf}
}
where $U(\Phi):=V(\phi(\Phi))$.

For $B=0$, there is no difference from the conventional Riemannian geometry, while for $B<0$, 
we always find $\Phi\leq \phi$.  In particular, $\phi$ will be constrained as
\eqn{
0\leq \phi\leq \frac{M_{\text{Pl}}}{\sqrt{|B|}}\,,
}
to avoid the ghost instability.  Therefore, the field value of $\phi$ cannot exceed the Planck mass typically in order that inflationary cosmology succeeds. 
 We found that the cases $\phi \ll M_{\rm pl}/\sqrt{|B|}$do not introduce new features for inflationary cosmology, so we will not discuss these cases furthermore.

When $B>0$, which only Models II(a), III(a) and III(c) admit, 
the redefined scalar field $\Phi$ behaves differently depending on its energy scale, such that
\beann
\Phi\approx  \left\{
\begin{array}{lc}
 \phi &(\phi\ll M_{\rm Pl}/\sqrt{B})  \\[.5em]
 {\sqrt{B}\over 2M_{\rm Pl}} \phi^2&(\phi\gg M_{\rm Pl}/\sqrt{B})   \\
\end{array} 
\right.
\,,
\enann
i.e., when $\phi$ is small, the difference between the metric-affine and its purely metric counterpart is relatively tiny, 
while the difference becomes significant when $\phi$ becomes larger than 
$M_{\rm Pl}/\sqrt{B}$.  In particular, during inflationary regime, the field value of $\phi$ can exceed $M_{\rm pl}$ in which the effective potential for the canonical field  $\Phi \propto \phi^2$ becomes flatter which may cause a smaller tensor-to-scalar ratio than the conventional scenario, $B=0$. 

The action (\ref{action_inf}) simply consists of a single canonical scalar field $\Phi$. 
To discuss an inflationary scenario, 
we analyze  the amplitude of the scalar perturbations as
\eqn{
P_\zeta \sim\frac{U}{24\pi^2 \epsilon_U}
\,,
}
and evaluate  the spectral index and tensor-to-scalar ratio as
\eqn{
n_s&\sim& 1+2\eta_U-6\epsilon_U\,,\\
r&\sim &16 \epsilon_U\,,
}
where
 the potential slow-roll parameters are defined by 
\eqn{
\epsilon_U(\Phi)&=&\frac{M_{\rm Pl}^2}2\l(\frac{U_{,\Phi}}{U}\r)^2,\\
\eta_U(\Phi)&=&M_{\rm Pl}^2\frac{U_{,\Phi\Phi}}{U}
\,.
}

Now assume the potential is a chaotic inflation type\cite{Linde:1983gd} as
\beann
V={1\over 2}m^2\phi^2\,,
\enann
and evaluate the observational parameters.
The modified potential is described as
\beann
U\approx  \left\{
\begin{array}{lc}
{1\over 2}m^2 \Phi^2 &(\Phi\ll M_{\rm Pl}/\sqrt{B})  \\[.5em]
{m^2M_{\rm Pl}\over \sqrt{B}}\, \Phi&(\Phi\gg M_{\rm Pl}/\sqrt{B})   \\
\end{array} 
\right.
\,.
\enann

If $B$ is very small, since the value of a scalar field is about the Planck mass at the end of inflation, 
we find the conventional chaotic inflation.
On the other hand, when $B$ is around the order of unity, the potential acts as a linear potential, which changes 
the inflationary scenario, and the observational parameters with it. \footnote{We chose $V(\phi)\propto \phi^2$ because of observational constraints. This could be taken as a general form of, for example, a polynomial  $V(\phi)\propto\phi^n$. In such case, the resulting effective potential will become $U(\Phi)\propto \Phi^n$ for the low energy case, and $U(\Phi)\propto \Phi^{\frac n2}$ for the high energy case. }

\begin{figure}[h]
    \centering\includegraphics[width=1.0\linewidth]{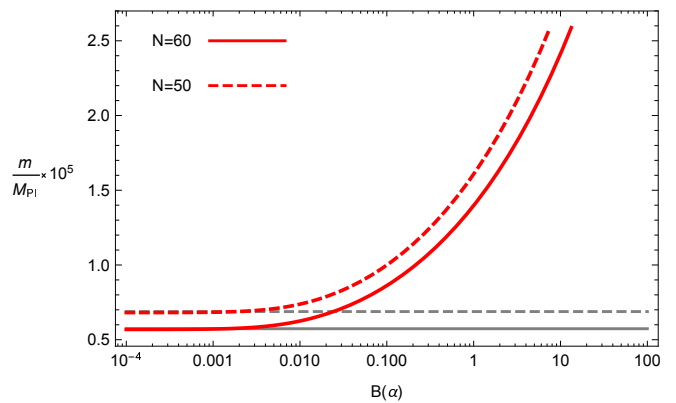}
    \caption{ The relation between the inflaton mass $m$ and the parameter $B(\alpha)$ 
    constrained from the observational amplitude of density fluctuations. The solid and dashed lines correspond to N=60 and 50, respectively.}
\label{Fig1}
 \end{figure}

Fig.\ref{Fig1} depicts the constraint on the inflaton mass $m$ and the parameter $B(\alpha)$ from the observed amplitude 
of the density fluctuation\cite{Aghanim:2018eyx}.
When the value of $B(\alpha)$ is small enough, we find the conventional chaotic inflationary model, which inflaton mass is fixed by the observation 
as $m\sim 6.45\times 10^{-6}M_{\rm Pl}$\cite{Aghanim:2018eyx}, while for $B(\alpha)\lsim 1$, the mass can be several times larger than the conventional model (see Fig.\ref{Fig1}).

\begin{figure}[h]
\includegraphics[width=.8\linewidth]{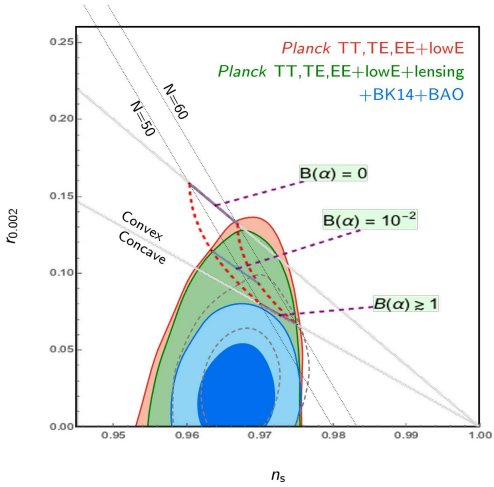}
    \caption{ The $n_s$-$r$ diagram for different values of $B(\alpha)$. 
The observational constraint is taken from Planck 2018\cite{Aghanim:2018eyx}.}
\label{Fig2}
\end{figure}

We also show the $n_s$-$r$ diagram in Fig. \ref{Fig2}.
From Fig.\ref{Fig2}, we find  that for a sufficiently small $B(\alpha)$ the potential acts as a chaotic potential, however, when $B(\alpha)$ is the order of unity, $r$ decreases. As a result, the model is not fully excluded from the current observations.
 For the upper bound of $r<0.10$, we find the constraint on the parameter $B$ as $B(\alpha)\gsim 0.034$ at $N=50$ e-folds. 
In words of the parameter $\alpha$, for the Models II(a) and III(a), we obtain 
\beann
\begin{array}{lclc}
\alpha\gsim 0.86&\text{or}&\alpha\lsim -0.13&\text{(Models II(a) and III(a))}
\\
\alpha\gsim 1.10 &\text{or}&\alpha\lsim -0.10&\text{(Model III(c))}\,.
  \\
\end{array} 
\enann
From the above result, 
if inflation was indeed caused by a chaotic inflation 
and the geometry was written in a metric-affine framework, one could say that
$\Gnabla\hskip -.3em{}^\mu\hskip -.4em\Gnabla_\mu$ ($\alpha=1$) is observationally 
favored than 
$\Gnabla_\mu\Gnabla\hskip -.3em{}^\mu$ ($\alpha=0$).

\subsection{G-inflation in metric-affine gravity}
\subsubsection{Scalar field with Galileon symmetry}

When we extend the kinetic terms of a scalar field, there exists one interesting approach, 
which is the so-called Horndeski 
scalar-tensor gravity theory, or its extended version\cite{Horndeski:1974wa,Kobayashi:2011nu,Tsujikawa:2014mba}.  The equation of motion in such theories
 consists of up to the second-order derivatives. Among such theories, the model with Galileon symmetry may be interesting because
it may be found in the decoupling limit of the DGP(Dvali-Gabadadze-Porrati) brane world model \cite{Dvali:2000hr,Nicolis:2004qq} (For reviews see for example \cite{Deffayet:2013lga}).\\
\indent The Galilean symmetry is defined by the transformation
such that 
\eqn{
\phi\to \phi +b^\mu x_\mu +c,
}
where $b^\mu$ and $c$ are some constants.

 In flat space, the Galileon symmetry 
fixes the Lagrangian of a scalar field as
\eqn{
\cf L_{(1,0)}&=&\phi\nn \\
\cf L_{(2,0)}&=&\d_\mu\phi\d^\mu\phi\nn \\
\cf L_{(3,0)}&=& \d_\mu\phi \d^\mu \phi \fsquare\phi-\d_\mu \phi \d_\nu\phi \d^\mu\d^\nu\phi 
\label{L301}\\
&=&\frac32\d_\mu\phi \d^\mu \phi \fsquare\phi +\text{(surface terms)}
\label{L302}
}
 up to cubic terms.
When we  covariantize the above terms, we have two starting points, i.e.,
 (\ref{L301}) and (\ref{L302}).
The covariantization $\fsquare\rightarrow \Gsquare$
contains one parameter $\alpha$ where
(see Eq. (\ref{Gsquare})), on the other hand
in the covariantization of the second term of  (\ref{L301}),
we have two possibilities;
$
\Gnabla\hskip -.3em{}_\mu \phi \Gnabla\hskip -.3em{}_\nu\phi \Gnabla\hskip -.3em{}^\mu\Gnabla\hskip -.3em{}^\nu\phi $ and 
$\Gnabla\hskip -.3em{}^\mu \phi \Gnabla\hskip -.3em{}^\nu\phi \Gnabla\hskip -.3em{}_\mu\Gnabla\hskip -.3em{}_\nu\phi
$, which may introduce one more parameter $\beta$.
Here just for simplicity, we analyze the first case.
The second case gives a similar result although we find the constraint on 
two parameters.

We shall analyze the following covariantized action 
\beann
S_{{\rm g} \phi}=\int \text{d}^4x \sqrt{-g} \l[\frac{M_{\rm Pl}^2}2\GR(g,\Gamma) -X
-\frac{X}{M^3}\Gsquare\phi\r]\,,
\enann
where $\displaystyle{X:=-{1\over 2}(\nabla\phi)^2}$ and $M$ is a parameter with mass dimension. This term is purely Galileon in flat space. 
Similar to the calculation in the previous section, the equivalent action in Riemannian geometry is obtained as
\bea
S_{{\rm g}\phi}&=&\int \text{d}^4x \sqrt{-g}\l[\frac{M_{\rm Pl}^2}2 \, R(g) \r.\nn\\
&&~~~~
\l.-X+\frac{4B(\alpha)}{M^2_{\rm Pl}M^6}X^3 -\frac{X}{M^3}\square\phi\r],\quad
\label{action_galilleon}
\ena
where $B(\alpha)$ is the same function of $\alpha$ as in the previous subsection \ref{chaotic_inflation}. 

\subsubsection{Emergence of G-inflation}
The action (\ref{action_galilleon}) is similar to the G-inflation action discussed in \cite{Kobayashi:2010cm},
where the non-linear term of $X$ naturally appears.
Note that the  third term  in our action is proportional to $X^3$  instead of  $X^2$ in the example proposed in \cite{Kobayashi:2010cm}, it also allows  de-Sitter solution as we will show it soon.

Assuming the flat FLRW spacetime, 
we find the Friedmann equations and the equation of motion of the scalar as
\begin{eqnarray*}
0&=&-3M^2_{\text{Pl}}H^2-\frac12\dot\phi^2+\frac3{M^3}H\dot\phi^3+\frac{5B}2\frac{\dot\phi^6}{M^6M^2_{\text{Pl}}},\\
0&=&M^2_{\text{Pl}}(3H^2+2\dot H)-\frac12\dot\phi^2+\frac{B}2\frac{\dot\phi^6}{M_{\text{Pl}}^2M^6}-\frac1{M^3}\ddot\phi\dot\phi^2,\\
0&=&\l(-1+3B\frac{\dot\phi^4}{M_{\text{Pl}}M^6}\r)(\ddot\phi+3H\dot\phi)+12B\ddot\phi\frac{\dot\phi^4}{M_{\text{Pl}}^2M^6}
\nn
\\
&&
+\frac{3}{M^3}\l(\dot H\dot\phi^2+2H\ddot\phi\dot\phi+3H^2\dot\phi^2\r)
\,,
\end{eqnarray*}
where $\displaystyle{H=\frac{\dot a}a}$ is the Hubble parameter defined by 
the scale factor $a(t)$.

In order to discuss an inflationary scenario, we first look for a de Sitter solution.
Assuming $H=H_{\text{dS}}$=constant and $\dot \phi=\dot \phi_{\text{dS}}$=constant,
we find two de-Sitter solutions as
\begin{eqnarray*}
X&=&X_{\text{dS}_\pm}:=
{M^3M_{\text{Pl}} \over \sqrt{3(1+4B)\pm \sqrt{3(3+16B)}}}
,\\
H&=&H_{\text{dS}_\pm}:=
{4M^3\over 3(1\pm \sqrt{3(3+16B)})\dot{\phi}_{\text{dS}_\pm}}
\,,
\end{eqnarray*}
where $X_{\text{dS}_\pm}=\dot{\phi}_{\text{dS}_\pm}^2/2$.

For the $+$ branch, $B>-3/16$ is required, while for the $-$ branch, we find $B>0$ or $-3/16<B<-1/6$.
Since $H$ must be positive, 
we find that $\dot{\phi}_{\text{dS}_+}$ is always positive while $\dot{\phi}_{\text{dS}_-}>0$ for $-3/16<B<-1/6$
and $\dot{\phi}_{\text{dS}_-}<0$ for $B>0$.
Models I, II (b) and III(b)  are ruled out because  $B=-3/8$ in the three models.


In order to study the stability of the de Sitter solution, we perturbed the present system.
The quadratic action of the scalar perturvation ${\cal R}_\phi$ within the unitary gauge ($\delta\phi=0$)
is obtained in  \cite{Kobayashi:2010cm,Kobayashi:2011nu}. In our case,  it becomes as
\beann
S^{(2)}_\pm&=&{\dot\phi_{\text{dS}_\pm}^2\over 2 (H_{\text{dS}_\pm}-\dot\phi_{\text{dS}_\pm}^3/2 M_{\rm Pl}^2M^3)^2}\\
&\times &\int d\eta d^3 x a^2
\left[G_S({\cal R}'_\phi)^2-F_S(\vect{\nabla}{\cal R}_\phi)^2\right]
\,,
\enann
where the prime denotes the differentiation with respect to the conformal time $\eta$, and 
\beann
F_S&=&{1\over 3}-{2(1+2B)X^2_{\text{dS}_\pm}\over M_{\rm Pl}^2 M^6}
\\
&=&
{8B\over 3\left[3+14B\pm(1+2B)\sqrt{3(3+16B)}\right]}\,,
\\
G_S&=&1+{6(1+6B)X^2_{\text{dS}_\pm}\over M_{\rm Pl}^2 M^6}
\\
&=&
{\left[3+16B\pm(1+2B)\sqrt{3(3+16B)}\right]\over 4B}\,.
\,.
\enann
If either $F_S$ or $G_S$ is negative, the de Sitter solution is unstable.
It is the case ($F_S<0$) when we choose the $-$ branch ($X_{\text{dS}_-}$ and $H_{\text{dS}_-}$).
While for the $+$ branch solution ($X_{\text{dS}_+}$ and $H_{\text{dS}_+}$),
 $G_S$ is laways positive for $B>-3/16$ but  $F_S$  becomes negative for $-3/16<B<0$.
As a result, one de Sitter solution ($X_{\text{dS}_+}$ and $H_{\text{dS}_+}$) is 
stable only when $B>0$, while the other solution is unstable.

\begin{figure}[h]
  \includegraphics[width=.9\linewidth]{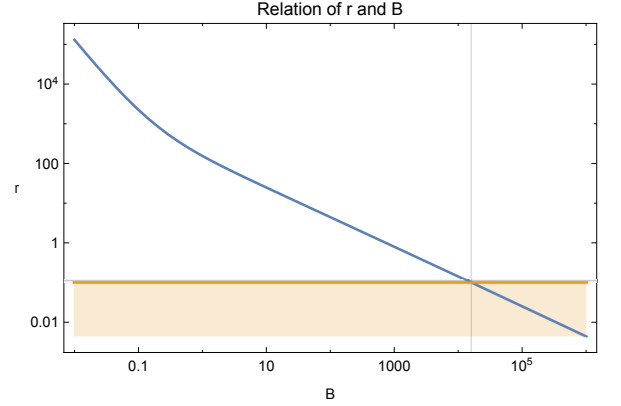}
    \caption{The tensor-to-scalar ratio $r$ in terms of  $B(\alpha)$ in metric-affine G-inflation.}
    \label{Fig3}
 \end{figure}

Once we know the solution of de Sitter phase, we can evaluate  the tensor-to-scalar ratio and the amplitude
of the scalar perturbations, 
which formula is given in \cite{Kobayashi:2010cm,Kobayashi:2011nu}, as
\begin{eqnarray*}
{\cal P}_\zeta 
&\sim&
\frac{B^2M^3}{27\pi^2M_{Pl}^3}
\sqrt{\frac{3+16B+\sqrt{3(3+16B})}{[2+11B+B\sqrt{3(3+16B)}]^3}}\,,
\\
r&\sim&\frac{6(1+6B)}{B^2}
\sqrt{\frac{6(1+6B)[1+\sqrt{3(3+16B)}]}{3+16B+\sqrt{3(3+16B)}}}\,.
\end{eqnarray*}
Since the observational upper bound of the tensor-to-scalar ratio is $r<0.10$\cite{Aghanim:2018eyx}, 
the constraint on $B$ becomes $B(\alpha) \gsim 1.6\times 10^4$ (see Fig. \ref{Fig3}  
, which 
corresponds to $\alpha \gsim 93.7$  or $\alpha \lsim -92.9$ for the Models II(a) and III(a),
 whereas  $\alpha\gsim  119.6$ or $\alpha\lsim-118.6$ for the Model III(c).   
 Furthermore, the mass parameter $M$ is also constrained from the amplitude of 
 the scalar perturbations and the constraint of $B(\alpha)$ as $M\lsim0.0060M_{\text{Pl}}$. 
 
Due to the shift symmetry of the action, 
one cannot end the de Sitter phase within this framework. 
The remedy, as in \cite{Kobayashi:2010cm}, is to introduce some function which 
breaks the scale invariance and to flip the sign of the ghost in the second term of the action. 
We have confirmed that the flip function whether polynomial or exponential can end inflation. 
In order to find a realistic inflation model, not only the de Sitter phase must end
but also  the spectral index $n_s$ 
should be tilted. 
Hence we may modify the action of the scalar field 
to 
\bea
S_{\phi}&=&\int \text{d}^4x \sqrt{-g} \l[ -g_1(\phi)X+g_2(\phi)
\frac{X}{M^3}\Gsquare\phi\r]\,,
\nonumber
\\
&=&
\int \text{d}^4x \sqrt{-g} \l[ -g_1(\phi)X+{4g_2^2(\phi)B\over M_{\rm Pl}^2 M^6}X^3
+{g_2(\phi)\over M^3}X\square\phi\r]\,.
\nonumber
\\
~
\label{G_action2}
\ena
where $g_1(\phi)$ and $g_2(\phi)$ are appropriate functions of $\phi$, which break the Galilean symmetry.
For example, we may choose $g_1(\phi)=\tanh[\lambda(\phi-\phi_{\text{end}})/M_{\rm Pl}]$ to finish an inflationary stage, 
while $g_2(\phi)=\exp[\epsilon_g \phi/M_{\rm Pl}]$ to title the density perturbations \cite{Kobayashi:2010cm}.
The parameter $\phi_{\rm end}$ can be absorbed  by redefinition of $M$.
In fact,  the action (\ref{G_action2}) is invariant for 
the shift transformation $\phi\rightarrow \tilde \phi=\phi-\phi_{\rm end}$
with the redefinition such that
\beann
M&\rightarrow& \tilde M=\exp\left(-{\epsilon_g  \phi_{\text{end}} \over 3 \lambda M_{\rm Pl}}\right)\,\lambda M\,,
\\
g_1(\phi)&\rightarrow& \tilde g_1(\tilde \phi)=\tanh \left( {\lambda\tilde \phi \over M_{\rm Pl}}\right)\,,
\\
g_2(\phi)&\rightarrow &\tilde g_2(\tilde \phi)=\exp\left({\epsilon_g\tilde \phi\over \lambda M_{\rm Pl}}\right)
\,.
\enann
Since $B$ is not changed, ${\cal P}_\zeta $ can be adjusted to the observational data by tuning 
the value of $\phi_{\rm end}$.

As for  the spectral index $n_s$, it is highly affected by the choice of those functions, 
thus we will not explicitly analyze here. 
We note that we can find an appropriate function to satisfy the observational data.

\section{Summary and Discussions}
 In this paper, assuming the Einstein-Hilbert action, we have classified the metric-affine theory of gravitation into three models (six cases). By separating the distortion tensor  from the connection, one can easily find the distortion tensor by solving an algebraic equation.  Since the connection is non-propagating, i.e. it does not have new degrees of freedom,  substituting the solution of the distortion tensor  into the metric-affine action, we  obtain
an equivalent effective action in the Riemannian geometry solely constructed by the metric, which differs from its counterpart model in Riemannian geometry.

If matter field does not couple to the connection (Palatini formalism), the effective action described in Riemannian geometry is equivalent to GR.
While if matter field couples to the  connection, an additional energy-momentum
appears from the coupling  in  the effective equations.
The additional terms by the distortion tensor are suppressed by  the Planck mass.
This Planck mass suppression is a characteristic feature that appears naturally in these metric-affine gravity theories,
and the additional terms will become important in a high-energy scale.

 We have then applied the formalism into two inflationary models: the `minimally' coupled model and the G-inflation type model. Both models are rather simple in the metric-affine case, and the models are characterized by the parameter $B(\alpha)$ which differs in the six classified cases, and the structure of the resultant action is all the same. 
In any case, the observational parameters are drastically different from the Riemannian geometry counterpart.
A key feature of minimally coupled models is that the effective potential becomes flatter than the conventional scenario in $\phi > M_{\rm pl}/\sqrt{B}$ with $B>0$ and then the tensor-to-scalar ratio becomes smaller. For instance, the chaotic inflation scenario is not fully excluded by the current observations in the metric-affine formalism. As for Galileon models, the metric-affine formalism naturally yields the $X^3$ term and a stable de Sitter solution, although G-inflationary model requires slightly unnatural large coupling parameters to be consistent with observations.  

Here we would like to comment on the 
 extension of the present  formalism.
Although we consider only the Einstein-Hilbert action (the scalar curvature) in this paper, 
 we can discuss more general action with the Riemann tensor or the higher-order terms of the curvatures.
In that case, we must  note that there are more curvature tensors than
 the usual Riemannian case since the Riemann tensor  does not satisfy some (anti-)symmetries (e.g. ${\GR}{}^\lambda_{~\mu\lambda\nu}\neq {\GR}{}_{\mu~\nu\lambda}^{~\lambda}, {\GR}{}^\lambda_{~\lambda\mu\nu}\neq0$).  
Nevertheless, we have confirmed that this is drastically simplified when
we assume a projective symmetry on the theory \cite{Aoki:2018lwx}. 
 The scalar-tensor theory of gravity in metric-affine geometry with projective invariance becomes equivalent to the DHOST theory, which guarantees that there is no ghost~\cite{Langlois:2015cwa,Crisostomi:2016czh,BenAchour:2016fzp}.
Hence metric-affine geometry could be a key to understanding ghost-free properties of scalar-tensor theories.\\\\

There are numerous extensions and applications that one may consider from this work. 
For example, one may first note that the properties of 'integrating out the connection' could only be done when the connection does not withhold new degrees of freedom. This is not the case of Palatini/metric-affine higher curvature gravity \cite{Borunda:2008kf,BasteroGil:2009cn,Dadhich:2010dg,Bernal:2016lhq,Exirifard:2007da}. When the higher order terms of the Riemannian curvatures are present, which may appear in the in quantum corrections, the connection obtains new degrees of freedom. The analysis is not simple, because the theory 
cannot be described by the effective action in Riemannian geometry. It is expected that metric-affine geometry differs greatly from Riemannian geometry. One may hope that metric-affine geometry will provide us rich phenomena that the Riemann case does not. 
  
 Another interesting issue that one must consider is that in metric-affine gravity, bosons and fermions react differently even in the standard model of particles\cite{Hehl:1974cn,Hehl:1976kj,Hehl:1994ue}.  This is due to the fact that the Dirac particles couple to both the metric and the connection, whereas gauge bosons just follow the orbit determined only by the metric. This is a key factor of metric-affine geometry since in principle all matter behaves alike in Riemannian geometry.  More specifically, geodesics of spin integers and spin halves will be different\cite{Audretsch:1981xn,Nomura:1991yx,Cembranos:2018ipn}. This may be able to be verified by tests of the equivalence principle\cite{Will:2014kxa,Will:2018bme}.  It would be also interesting to see if whether there could be imprints of metric-affine geometry through the CMB(bosons) and the cosmic neutrino background  [C$\nu$B] (fermions), if any, which could lead to verification from future observations.\cite{Aghanim:2018eyx,Betts:2013uya,Tanabashi:2018oca,Baracchini:2018wwj} \\
\indent Finally, we would also  like to mention on further application to cosmology.  Recently, interesting results appeared from the Palatini approach of Higgs inflation \cite{Bauer:2008zj,Bauer:2010jg,Rasanen:2017ivk}.  It would be interesting to extend this to further cases such as New Higgs inflation\cite{Germani:2010gm} and Hybrid Higgs inflation\cite{Sato:2017qau}. In addition, since fermions and bosons react differently in the metric-affine formalism, it is hoped that the reheating phase would compute different results. This fact, that fermions and bosons couple differently with the Higgs, has not been considered in any literature and it is worth investigating using a concrete model such as above. 

\section*{Acknowledgements}
K.S. would like to Shoichiro Miyashita, Masahide Yamaguchi and Shinji Mukohyama  for eye-opening discussions and advice throughout this work. We would also thank Tommi Tenkanen for fruitful comments. Furthermore, K.S. will like to thank the Yukawa Institute for Theoretical Physics for their kind hospitality during his stay. The work of K.A. was supported in part by a Waseda University Grant for Special Research Projects (No. 2018S-128). This work was also supported in part by JSPS KAKENHI Grant Numbers JP16K05362 (KM)  
and JP17H06359 (KM). 

\newpage

\appendix
\begin{widetext}
\section{Metric-affine General Relativity rewritten with Torsion or Non-metricity}
As shown in Section 2 distortion tensor and the Riemann tensor could be written as
\eqn{
{\kappa^\lambda}_{\mu\nu}&=&\frac12\l({\cf T^\lambda}_{\mu\nu}+{{\cf T_\nu}^\lambda}_\mu-{\cf T_{\mu\nu}}^\lambda +{{\cf Q_\nu}^\lambda}_\mu+{\cf Q_{\mu\nu}}^\lambda-{\cf Q^\lambda}_{\mu\nu}\r)\,,\\
{{\GR}{}^\alpha}_{\beta\mu\nu}(\Gamma)&=&{R^\alpha}_{\beta\mu\nu}(\{\})+\nabla_\mu {\kappa^\alpha}_{\nu\beta}-\nabla_\nu {\kappa^\alpha}_{\mu\beta}+{\kappa^\alpha}_{\mu\lambda}{\kappa^\lambda}_{\nu\beta}-{\kappa^\alpha}_{\nu\lambda}{\kappa^\lambda}_{\mu\beta}\,.
}
In this section, we will write the explicit form of equation of motions in General Relativity, when either the torsionless or metric-compatiblility is satisfied, with using the torsion tensor and the non-metricity tensor. 
\subsection{Metric-Affine EH}
\subsubsection{Torsionless}
For $\cf T=0$ and $\cf L_g={\GR}(g,\Gamma)$ the Einstein Hilbert could be rewritten as,
\eqn{
{\GR}{}(g,\Gamma)\=R(g,\{\})+\-4 \cf Q_{\lambda\mu\nu}\cf Q^{\lambda\mu\nu}-\-2 \cf Q_{\lambda\mu\nu}\cf Q^{\mu\lambda\nu}+2 {\cf Q}_\mu \cf W^\mu -4 \cf W^\mu \cf W_\mu\,.
}
The eom for the connection could be derived by taking caution of the symmetry of the connection (the last two indices are symmetric) as, 
\eqn{
\-{\sqrt{-g}}\frac{\delta S_g}{\delta {\kappa^\alpha}_{\beta^\prime\gamma^\prime}}\times \frac12\l(\delta^\beta_{\beta^\prime}\delta^\gamma_{\gamma^\prime}+\delta^\gamma_{\beta^\prime}\delta^\beta_{\gamma^\prime}\r)\=-{\cf Q_\alpha}^{\beta\gamma}+2\cf Q_\alpha g^{\beta\gamma}-2\cf W^{(\beta}\delta^{\gamma)}_\alpha + {\cf Q}^{(\beta}\delta^{\gamma)}_\alpha\,.
}
If you take the variation of the action wrt to $\cf Q$,
\eqn{
\-{\sqrt{-g}}\frac{\delta S_g}{\delta {\cf Q_\mu}^{\nu\lambda}}\=\frac12{\cf Q^\mu}_{\nu\lambda}-\frac14{\cf Q_{(\lambda\nu)}}^\mu +\frac12g_{\nu\lambda} {\cf Q}^\mu+2\delta^\mu_{(\nu}\cf W_{\lambda)}-2g_{\nu\lambda}\cf W^\mu\,.
}
Notice that there is an equality
\eqn{
\-{\sqrt{-g}}\frac{\delta S_g}{\delta {\cf Q_\mu}^{\nu\lambda}}\times\l(\delta_\mu^\beta g^{\gamma(\nu}+\delta_\mu^\gamma g^{\beta(\nu}\r)\delta_\alpha^{\lambda)}=
\-{\sqrt{-g}}\frac{\delta S_g}{\delta {\kappa^\alpha}_{\beta^\prime\gamma^\prime}}\times \frac12\l(\delta^\beta_{\beta^\prime}\delta^\gamma_{\gamma^\prime}+\delta^\gamma_{\beta^\prime}\delta^\beta_{\gamma^\prime}\r)\,.
}
Since
\eqm{
\frac{\delta {\cf Q_\mu}^{\nu\lambda}}{\delta {\kappa^\alpha}_{\beta^\prime\gamma^\prime}} \frac12\l(\delta^\beta_{\beta^\prime}\delta^\gamma_{\gamma^\prime}+\delta^\gamma_{\beta^\prime}\delta^\beta_{\gamma^\prime}\r)\=\delta^{\beta^\prime}_\mu \delta^{(\lambda}_\alpha g^{\nu)\gamma^\prime} \l(\delta^\beta_{\beta^\prime}\delta^\gamma_{\gamma^\prime}+\delta^\gamma_{\beta^\prime}\delta^\beta_{\gamma^\prime}\r)\\
\=\l(\delta^{\beta}_\mu  g^{\gamma(\nu}+\delta^{\gamma}_\mu  g^{\beta(\nu}\r)\delta^{\lambda)}_\alpha\,,
}
this should also hold for the matter sector with its hyper energy-momentum tensor as,
\eqm{
\frac{\delta S_M}{\delta {\cf Q_\mu}^{\nu\lambda}}\times\l(\delta_\mu^\beta g^{\gamma(\nu}+\delta_\mu^\gamma g^{\beta(\nu}\r)\delta_\alpha^{\lambda)}=
\frac{\delta S_M}{\delta {\kappa^\alpha}_{\beta^\prime\gamma^\prime}}\,.
}
The equation for the metric is
\eqn{
\-{\sqrt{-g}}\frac{\delta S_g}{\delta g^{\mu\nu}}\={\GR}{}_{(\mu\nu)}(\Gamma)-\frac12g_{\mu\nu}{\GR}{}(g,\Gamma)\nn\\
\=G_{\mu\nu}(g,\{\})+2g_{\mu\nu}\cf Q^\lambda \cf Q_\lambda -g_{\mu\nu}\cf W^\mu { \cf  Q}_\mu +2\cf W^\alpha \cf Q_{(\mu\nu)\alpha}-\cf  W^\alpha\cf  Q_{\alpha\mu\nu} \nn\\
&&+\-4 g_{\mu\nu}\cf  Q^{\alpha\beta\gamma}\cf Q_{\beta\alpha\gamma} -\frac18 g_{\mu\nu} \cf Q^{\alpha\beta\gamma}\cf Q_{\alpha\beta\gamma}+\frac12 {\cf Q^{\alpha\beta}}_\mu \cf Q_{\alpha\beta\nu}-\-2 {\cf Q^{\alpha\beta}}_\mu \cf Q_{\beta\alpha\nu}-\-4 \cf Q_{\mu\alpha\beta}{\cf Q_\nu}^{\alpha\beta}\,.\nn\\
}

\subsubsection{Metric Compatible}
For $\cf Q=0$, the following holds as, 
\eqn{
{\GR}{}(g,\Gamma)\=R(g,\{\})-\cf T^\mu\cf  T_\mu+\-4\cf T_{\lambda\mu\nu}\cf T^{\lambda\mu\nu}+\frac12\cf T_{\lambda\mu\nu}\cf T^{\mu\lambda\nu}\,.
}
The EoM for the connection could be derived by taking caution of the symmetry of the connection (the last first and third indices are anti-symmetric) as, 
\eqn{
\-{\sqrt{-g}}\frac{\delta S_g}{\delta {\kappa^{\alpha^\prime}}_{\beta\gamma^\prime}}\times\frac12\l(\delta_\alpha^{\alpha^\prime}\delta^\gamma_{\gamma^\prime}-g_{\alpha\gamma^\prime}g^{\alpha^\prime\gamma}\r)\=\cf T^{\beta\ \gamma}_{\ \alpha}-g^{\beta\gamma}\cf T_\alpha +\delta^\beta_\alpha\cf  T^\gamma\,.
}
If you take the action w.r.t. to the torsion,
\eqn{
\frac{1}{\sqrt{-g}}\frac{\delta S_g}{\delta {\cf T^\lambda}_{\mu\nu}}={\cf T_\lambda}^{\mu\nu}-{\cf T^{[\mu\nu]}}_\lambda+\delta^{[\mu}_\lambda\cf  T^{\nu]}\,,
}
Now since,
\eqn{
\frac{1}{\sqrt{-g}}\frac{\delta S_g}{\delta {\cf T^\lambda}_{\mu\nu}}\times\delta^\beta_{[\mu}\l(\delta^\gamma_{\nu]}\delta^\lambda_\alpha-g_{\nu]\alpha}g^{\lambda\gamma}\r)=\-{\sqrt{-g}}\frac{\delta S_g}{\delta {\kappa^{\alpha^\prime}}_{\beta\gamma^\prime}}\times\frac12\l(\delta_\alpha^{\alpha^\prime}\delta^\gamma_{\gamma^\prime}-g_{\alpha\gamma^\prime}g^{\alpha^\prime\gamma}\r)\,.
}
The equation for the metric is
\eqn{
\-{\sqrt{-g}}\frac{\delta S_g}{\delta g^{\mu\nu}}\={\GR}{}_{(\mu\nu)}(\Gamma)-\frac12g_{\mu\nu}{\GR}{}(g,\Gamma)\nn\\
\=G_{\mu\nu}(g,\{\})+\frac12 g_{\mu\nu}\cf T^\alpha \cf T_\alpha+\-2\cf T_\lambda {\cf T_{(\mu\nu)}}^\lambda\nn\\
&&-\frac18 g_{\mu\nu} \cf T_{\alpha\beta\gamma}\cf T^{\alpha\beta\gamma}-\frac14 g_{\mu\nu} \cf T_{\alpha\beta\gamma}\cf T^{\beta\alpha\gamma}-\frac14\cf T_{\alpha\beta(\mu}{\cf T_{\nu)}}^{\alpha\beta} +\frac14\cf T_{\mu\alpha\beta}{\cf T_\nu}^{\alpha\beta}\,.
}

\section{An extended d'Alembertian and the constraints from the observational data for inflation}
Here we consider an extended version of d'Alembertian (\ref{Gsquare_alpha}), which is
\bea
\Gsquare\phi:=\square\phi +\left(\alpha_{\cf Q}{\cal Q}^\lambda+\alpha_{\cf W}{\cal W}^\lambda+
\alpha_{\cf T}{\cal T}^\lambda\right)\partial_\lambda \phi
\,.
\label{extended_square}
\ena
where
$
{\cal T}_\mu =T^\lambda_{~\mu\lambda}\,,
{\cal  W}_\mu =\frac14 Q_{\mu\lambda}^{~~\lambda}\,,
 {\cal Q}_\mu = Q^\lambda_{~\lambda\mu}$. With the choice of $\alpha_{\cf Q}=-\alpha,\alpha_{\cf W}=2,\alpha_{\cf T}=-1$ the d'Alembertian reduces to (\ref{Gsquare_alpha}). 
We assume a canonical scalar field which 
action is given by (\ref{eq:action_phi}), and then 
 present the Riemann effective action
 \bea
S_{{\rm g}\phi}&=&\int d^4x\sqrt{-g} \l[\frac{M_{\text{Pl}}^2}2R(g)
-{1\over 2}\left(1+{B(\alpha_I)\over M_{\rm Pl}^2}\phi^2\right)(\nabla\phi)^2-V(\phi)\r]
\,,
\label{effective_action_A}
\ena
in each models discussed in \S. \ref{scalar_field_metric_affine}.
Based on this reduction, we show the constraints on the parameters 
$\alpha_I$ ($I=Q$,$W$, and $T)$ from the observational data for inflation.
Note that in the further calculations, all of the coefficients $\alpha_I$  can be also 
arbitrary functions 
of $\phi $ and $X$ as $\alpha_I(\phi,X)$, however, just for simplicity, the analysis would be done assuming $\alpha_I$ as a constant.  Furthermore, this could be considered as a scalar-tensor theory minimally coupled to the connection. A similar action was considered in the context of classification of torsionless metric-affine scalar-tensor theories through the transformation of the metric and the connection in \cite{Kozak:2017yez,Kozak:2018vlp}.

\subsection{Model I (Projective Invariant Model)}
The projective transformation (\ref{projective_transformation}) gives 
\beann
 \Gsquare\phi \rightarrow  \tGsquare\phi &=&\square\phi +\alpha_{\cf T} (\cf
 T^\mu+3U^\mu) \d_\mu\phi+\alpha_{\cf W}( \cf W^\mu+2U^\mu) \d_\mu\phi
 +\alpha_{Q} ({\cf  Q}^\mu+2U^\mu) \d_\mu\phi\nn\\
&=&\Gsquare \phi +[3\alpha_{T}+2(\alpha_{W}+\alpha_{Q})]U^\mu \d_\mu\phi
\,.
\enann
Hence, in order for the theory to have projective invariance, one must impose the condition of 
 $3\alpha_{T}+2(\alpha_{W}+\alpha_{Q})=0$.

Furthermore we find the following solution:
\eqn{
\bar \kappa{^\alpha}_{\beta\gamma}
 &= &\frac{\phi}{4M_{Pl}^2}\l[(3\alpha_{\cf T}+\alpha_{\cf W})g_{\beta\gamma}\d^\alpha\phi +(\alpha_{\cf T}+\alpha_{\cf W})\delta^\alpha_\beta \d_\gamma\phi\r]\,,\\
{\cf T^\lambda}_{\mu\nu}&=&\frac{\alpha_{\cf T}+\alpha_{\cf W}}{4M_{Pl}^2}\phi\delta^\lambda_{[\mu}\d_{\nu]}\phi\,,
\\
{\cf Q_\lambda}^{\mu\nu}&=&\frac{2\alpha_{\cf T}+\alpha_{\cf W}}{M^2_{Pl}} \phi\delta^{(\mu}_\lambda\d^{\nu)}\phi
\,,
}
up to gauge freedom.
As a result, the Riemann equivalent action becomes (\ref{effective_action_A}) 
with  
\eqn{
B(\alpha_I)
=-\frac18\l(27\alpha_{\cf T}^2+11\alpha_{\cf W}^2+34\alpha_{\cf T}\alpha_{\cf W}+20\alpha_{\cf W}\alpha_{\cf Q}+40\alpha_{\cf T}\alpha_{\cf Q}\r)
\,.\label{extMod1}
}

\subsection{Model II}
\subsubsection{\rm Model II(a) (Torsion-free Model)}
Assuming ${\cf T^\mu}_{\nu\rho}=0$ (and thus $\alpha_{\cf T}=0$), we find 
\eqn{
{\kappa^\alpha}_{\beta\gamma}\=-\frac{\phi}{12M^2_{Pl}}\l[3(\alpha_{W}+2\alpha_{Q})g_{\beta\gamma} \d^\alpha \phi -2(\alpha_{W}-2\alpha_{Q})\delta^\alpha_{(\beta}\d_{\gamma)}\phi \r]\,,\\
{\cf Q_\lambda}^{\mu\nu}\=\frac{\phi}{6M^2_{Pl}}\l[(\alpha_{\cf W}-2\alpha_{\cf Q})g^{\mu\nu}\d_\lambda\phi -2(\alpha_{\cf W}+4\alpha_{\cf Q})\delta_\lambda^{(\mu}\d^{\nu)}\phi\r]
\,.
}

The Riemann equivalent action becomes (\ref{effective_action_A}) with 
\eqn{
B(\alpha_I)= -\frac{\alpha^2_{W}-16\alpha_{W}\alpha_{Q}-44\alpha^2_{Q}}{24} \label{extMod2a}
\,.
}

\subsubsection{\rm Model II(b) (Metric-Compatible Model)}
Assuming $\cf Q=0$ ($\alpha_{\cf W}=\alpha_{\cf Q}=0$), we find
\eqn{
{\kappa^\alpha}_{\beta\gamma}\=\frac{\alpha_{\cf T}}{4M^2_{Pl}}\phi\l(g_{\beta\gamma}\d^\alpha\phi-\delta^\alpha_\beta\d_\gamma\phi \r)\,,\\
{\cf T^\alpha}_{\beta\gamma}\=-\frac{\alpha_{\cf T}}{2M^2_{Pl}}\phi\delta^\alpha_{[\beta}\d_{\gamma]}\phi
\,.
}

\end{widetext}
The Riemann equivalent action becomes (\ref{effective_action_A})  with 
\eqn{
B(\alpha_I)=-\frac{3}{8}\alpha_{\cf T}^2
\,.
}

\subsection{Model III (constraint with a Lagrange Multiplier)}
In the Model III, the Lagrange Multiplier $\lambda_\mu$ is introduces 
to fix the gauge freedom.
We find the following solutions for each model.

\subsubsection{\rm Model III(a) (${\cal T}_\mu=0, \alpha_{\cf T}=0$)}
The solution is 
\beann
{\kappa^\alpha}_{\beta\gamma}&=&-\frac{\phi}{12M^2_{Pl}}\Big[3(\alpha_{W}+2\alpha_{Q})g_{\beta\gamma} \d^\alpha \phi \\
&&
~~~~~~~~~~~~-2(\alpha_{W}-2\alpha_{Q})\delta^\alpha_{(\beta}\d_{\gamma)}\phi \Big]\,,\\
{\cf T^\alpha}_{\beta\gamma}&=&0\,,\\
{\cf Q_\lambda}^{\mu\nu}&=&\frac{\phi}{6M^2_{Pl}}\l[(\alpha_{\cf W}-2\alpha_{\cf Q})g^{\mu\nu}\d_\lambda\phi -2(\alpha_{\cf W}+4\alpha_{\cf Q})\delta_\lambda^{(\mu}\d^{\nu)}\phi\r]\,,
\enann
with 
\beann
\lambda^\mu&=&\lambda_T^\mu \equiv -\frac2{3M_{Pl}^2} (\alpha_{\cf W}+\alpha_{\cf Q})\phi\d^\mu\phi\,.\\
\enann
The Riemann equivalent action becomes (\ref{effective_action_A}) with 
\eqn{
B(\alpha_I)=-\frac{\alpha^2_{W}-16\alpha_{W}\alpha_{Q}-44\alpha^2_{Q}}{24}\label{extMod3a}
\,.
}


\subsubsection{\rm Model.III(b) ($ {\cf Q}_\mu=0, \alpha_{\cf Q}=0$)}
The solution is 
\beann
{\kappa^\alpha}_{\beta\gamma}\=\frac{3C_T+C_W}{4M^2_{Pl}}\phi g_{\beta\gamma}\d^\alpha\phi+\frac{\alpha_{\cf T}+\alpha_{\cf W}}{4M^2_{Pl}}\phi \delta^\alpha_\beta\phi\d_\gamma
\\
&&
-\frac{5(\alpha_{\cf T}+2\alpha_{\cf W})}{4M^2_{Pl}}\phi \delta^\alpha_\gamma\phi\d_\beta\,,\\
{\cf T^\alpha}_{\beta\gamma}\=\frac{11\alpha_{\cf T}+6\alpha_{\cf W}}{8M^2_{Pl}}\phi\delta^\alpha_{[\beta}\d_{\gamma]}\phi\,,\\
{\cf Q_\alpha}^{\beta\gamma}\=-\frac{2\alpha_{\cf T}+\alpha_{\cf W}}{2M^2_{Pl}}\phi\l(5g^{\beta\gamma}\d_\alpha\phi -2\delta_\alpha^{(\beta}\d^{\gamma)}\phi\r)\,,
\enann
with
\beann
\lambda^\mu=\lambda_W^\mu
\equiv -\frac1{2M_{Pl}^2} (3\alpha_{\cf T}+2\alpha_{\cf W})\phi\d^\mu\phi\,.
\enann
The Riemann equivalent action becomes (\ref{effective_action_A}) with 
\beann
B(\alpha_I)=\frac{3}{8}(11\alpha_{\cf T}^2+12\alpha_{\cf T}\alpha_{\cf W}+3\alpha_{\cf W}^2) \label{extMod3b}
\,.
\enann



\subsubsection{\rm Model III(c) (${\cal W}_\mu=0, \alpha_{\cf W}=0$)}

The solution is 
\beann
{\kappa^\alpha}_{\beta\gamma}&=&\frac{3\alpha_{\cf T}-2\alpha_{\cf Q}}{8M^2_{Pl}}\phi g_{\beta\gamma}\d^\alpha\phi-\frac{\alpha_{\cf T}+2\alpha_{\cf Q}}{8M^2_{Pl}}\phi \delta^\alpha_\beta\phi\d_\gamma
\nonumber 
\\
&&-\frac{\alpha_{\cf T}-2\alpha_{\cf Q}}{16M^2_{Pl}}\phi \delta^\alpha_\gamma\phi\d_\beta\,,
\\
{\cf T^\alpha}_{\beta\gamma}&=&-\frac{\alpha_{\cf T}+6\alpha_{\cf Q}}{8M^2_{Pl}}\phi\delta^\alpha_{[\beta}\d_{\gamma]}\phi\,,
\\
{\cf Q_\alpha}^{\beta\gamma}&=&-\frac{\alpha_{\cf T}-2\alpha_{\cf Q}}{8M^2_{Pl}}\phi\l(g^{\beta\gamma}\d_\alpha\phi -4\delta_\alpha^{(\beta}\d^{\gamma)}\phi\r)\,,\\
\enann
with 
\beann
\lambda^\mu\=\lambda_Q^\mu\equiv -\frac1{M_{Pl}^2} (3\alpha_{\cf T}-2\alpha_{\cf Q})\phi\d^\mu\phi\,.
\enann
The Riemann equivalent action becomes (\ref{effective_action_A}) with 
\beann
B(\alpha_I)=-\frac3{32}(\alpha_{\cf T}^2+12\alpha_{\cf T}\alpha_{\cf Q}-12\alpha_{\cf Q}^2)
\,.\label{extMod3c}
\enann


\subsection{Relation between the three models}

We find one interesting result, which is some relation between the three models I, II, and III.

In Model I, since the theory is invariant under the projective transformation,
we can eliminate one of the three connection terms using the gauge freedom.
For example, when we choose 
$U^\mu=-{\cal T}^\mu/3$, the connection term of ${\cal T}^\mu$ disappears.
Then only two parameters $\alpha_{\cal Q}$ and $\alpha_{\cal W}$ remain 
in the extended d' Alembertian (\ref{extended_square}).
Similarly when we choose $U^\mu=-{\cal W}^\mu/2$ and $U^\mu=- {\cal Q}^\mu/2$, 
we find two parameters $\alpha_{\cal T}, \alpha_{\cal Q}$ and $\alpha_{\cal T}, \alpha_{\cal W}$ in  (\ref{extended_square}), respectively.

The solution for each case  becomes
\beann
&&
\bar \kappa{^\alpha}_{\beta\gamma}
\\
&&~~=
  \left\{
    \begin{array}{l}
     \displaystyle{\frac{\phi}{12M^2_{\rm Pl}}\l[-3(\alpha_{W}+2\alpha_{Q})g_{\beta\gamma} \d^\alpha \phi 
+(\alpha_{W}-2\alpha_{Q})\delta^\alpha_\beta\d_\gamma\phi \r] } \\[1em]
   \displaystyle{ \frac{\phi}{8M^2_{\rm Pl}}\l[(3\alpha_{T}-2\alpha_{Q})g_{\beta\gamma} \d^\alpha \phi -(\alpha_{T}+2\alpha_{Q})\delta^\alpha_\beta\d_\gamma\phi \r] }  \\[1em]
    \displaystyle{   \frac{\phi}{4M^2_{\rm Pl}}\l[(3\alpha_{T}+\alpha_{W})g_{\beta\gamma} \d^\alpha \phi +(\alpha_{T}+\alpha_{W})\delta^\alpha_\beta\d_\gamma\phi \r]} 
    \end{array}
  \right.
\,,
\enann
with the gauge function $U^\mu$ as
\beann
U_\mu
= \left\{
    \begin{array}{l}
 \displaystyle{\frac{\alpha_{\cf W}-2\alpha_{\cf Q}}{12M^2_{Pl}}\phi\d_\mu\phi}
\\[1em]
 \displaystyle{-\frac{\alpha_{\cf T}-2\alpha_{\cf Q}}{16M^2_{Pl}}\phi\d_\mu\phi}
\\[1em]
 \displaystyle{-\frac{5(2\alpha_{\cf T}+\alpha_{\cf W})}{4M^2_{Pl}}\phi\d_\mu\phi}
    \end{array}
  \right.
\,.
\enann
The parameter $B(\alpha_I)$ becomes
\beann
B(\alpha_I)
= \left\{
    \begin{array}{l}
 \displaystyle{-\frac{1}{24}(\alpha^2_{W}-16\alpha_{W}\alpha_{Q}-44\alpha^2_{Q})}
\\[1em]
 \displaystyle{ -\frac{3}{32}(\alpha_{\cf T}^2+12\alpha_{\cf T}\alpha_{\cf Q}-12\alpha_{\cf Q}^2)}
\\[1em]
 \displaystyle{\frac{3}{8}(11\alpha_{\cf T}^2+12\alpha_{\cf T}\alpha_{\cf W}+3\alpha_{\cf W}^2)}
    \end{array}
  \right.
\,,
\label{B_modelI}
\enann
which are obtained from (\ref{extMod1}) by eliminating one parameter by use of the projective invariance condition $3\alpha_{\cal T}+2(\alpha_{\cal W}+\alpha_{\cal Q})=0$.

When we compare the above  results with those in Models II or III, 
we find that 
Models III (a), (b) and (c) correspond to the above three values in Model I (\ref{B_modelI}), respectively. 
We also find that Model II (a) is the same as the first case in Model I (\ref{B_modelI}).
As for Model II (b), it cannot be obtained from Model I with an appropriate gauge choice 
except for some special case of the parameters.  Only the case of 
 $\alpha_{\cal W}=-2\alpha_{\cal T}$ and $\alpha_{\cal Q}=\frac12 \alpha_{\cal T}$,
 which satisfies the projective invariance condition,  
we find the same result for Model II (b) and Model I.

Although the effective action in Model III is the same as that in the Model I with gauge fixing,
the Model III has no projective invariance in general.
If we impose the projective invariant condition for three parameters in Model III, 
we expect that it is a subclass of the Model I with the corresponding gauge fixing.

\subsection{Observational constraints on the parameters $\alpha_I$}
\begin{figure}[h]
  \includegraphics[width=.5\linewidth]{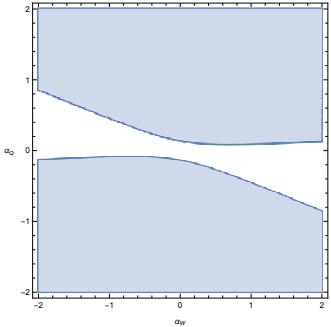}
    \caption{Constraints on $\alpha_W$ and $\alpha_Q$ in Model I with two 
    parameters ($\alpha_W, \alpha_Q$), and Model II(a) and Model III(a).
    The shaded region is consistent with the observational data for the tensor-scalar ratio $r$.}
    \label{FigA61}
 \end{figure}

\begin{figure}[h]
  \includegraphics[width=.5\linewidth]{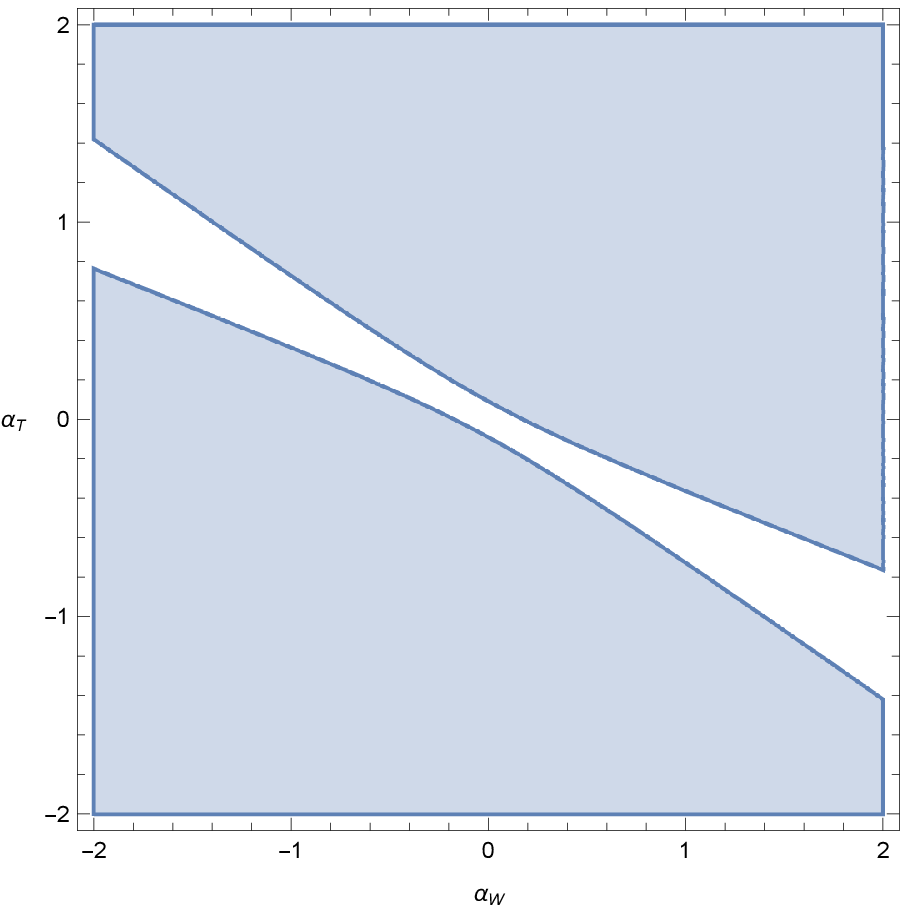}
    \caption{Constraints on $\alpha_T$ and $\alpha_W$ in Model I with two 
    parameters ($\alpha_T,\alpha_W$), and  Model III(b).
    The shaded region is consistent with the observational data for the tensor-scalar ratio $r$.}
    \label{FigA62}
 \end{figure}

\begin{figure}[h]
  \includegraphics[width=.5\linewidth]{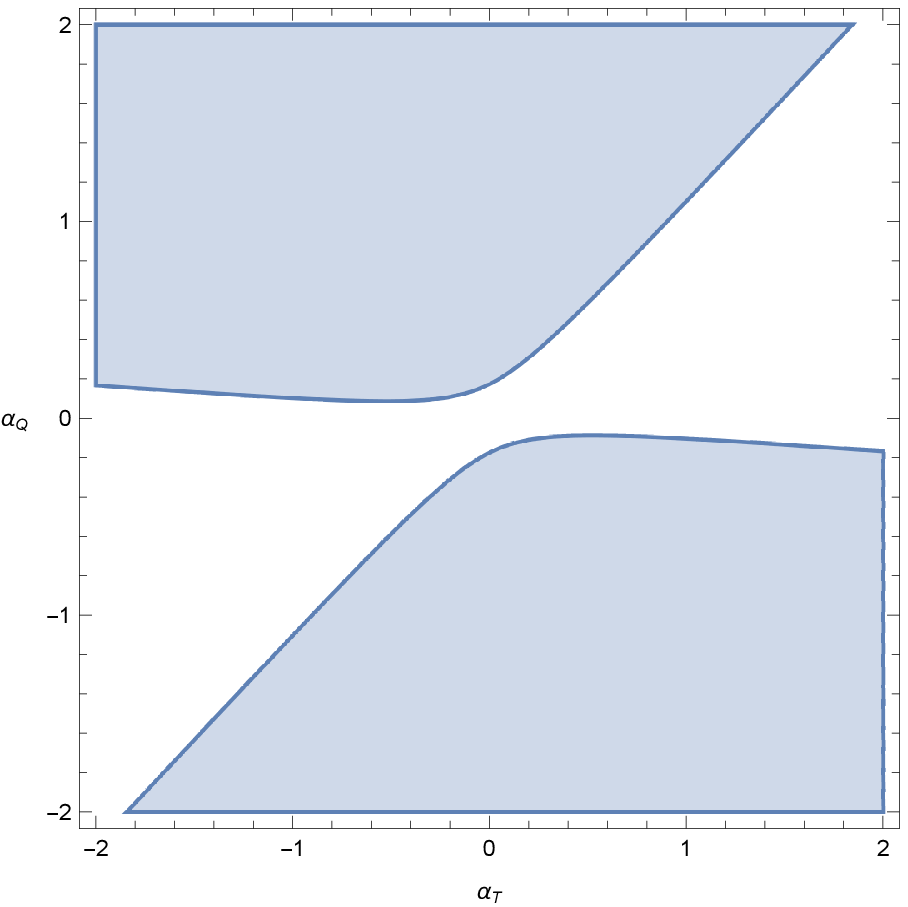}
    \caption{Constraints on $\alpha_Q$ and $\alpha_T$ in Model I with two 
    parameters ($\alpha_Q,\alpha_T$), and  Model III(c).
    The shaded region is consistent with the observational data for the tensor-scalar ratio $r$.}
    \label{FigA63}
 \end{figure}

In this final section, we consider the  observational constraints on
 the parameters in the extended d'Alembertian
 for the chaotic potential $V(\phi)=\frac12m^2\phi^2$.
From the observational constraints on the tensor-mass ratio $r$, we find 
 $B(\alpha_I)\gsim 0.034$. 
The projective invariant Model I consists of three parameters,
which must satisfy the projective condition $3\alpha_T+2(\alpha_W+\alpha_Q)=0$.
By projecting the observational constraints on $B(\alpha_I)$ with   (\ref{extMod1}) 
onto  two-parameter plane,
we find the allowed regions for two parameters 
shown  in Figs. \ref{FigA61}, \ref{FigA62} and  \ref{FigA63}.

Models  III(a), III(b) and III(c) give the same function 
 $B(\alpha_I)$ as those in Model I with specific gauges (${\cal T}^\mu=0$), 
 (${\cal Q}_\mu=0$) and (${\cal W}_\mu=0$), respectively.
As a result, the constraints on the two parameters are  given by Figs. \ref{FigA61}, \ref{FigA62} and  \ref{FigA63}, respectively.
Model II(a) is the same as Model III(a), which constraints on two parameters are 
shown in Fig. \ref{FigA61}.
Model II(b) is observationally excluded because of $B(\alpha_I)<0$.

\newpage

\bibliography{Inflation_metric_affine_ref}

\end{document}